\documentclass[longauth]{aa}  
\usepackage[hidelinks,colorlinks=true,linkcolor=blue,citecolor=blue]{hyperref}
\usepackage{graphicx}
\usepackage{txfonts}
\usepackage{amsmath}
\usepackage{xcolor}
\usepackage{euclid}
\usepackage{natbib}
\def\code#1{\texttt{#1}}

\newcommand{\Mpch}{\(h^{-1}\,\mathrm{Mpc}\;\)}

\defcitealias{tinker}{T10}
\defcitealias{castro23}{C23}
\defcitealias{2024AA...691A..62E}{C24}

\begin{document}

   \title{\Euclid preparation}
\subtitle{CX. Testing template-fitting models for the multipoles of the two-point clustering of galaxy clusters
}    

   \titlerunning{Cluster clustering multipoles}

\newcommand{\orcid}[1]{} 	   
\author{Euclid Collaboration: E.~Tsaprazi\orcid{0000-0001-5082-4380}\thanks{\email{e.tsaprazi@imperial.ac.uk}}\inst{\ref{aff1}}
\and A.~Fumagalli\orcid{0009-0004-0300-2535}\inst{\ref{aff2}}
\and F.~Marulli\orcid{0000-0002-8850-0303}\inst{\ref{aff3},\ref{aff4},\ref{aff5}}
\and A.~Heavens\orcid{0000-0003-1586-2773}\inst{\ref{aff1}}
\and G.~F.~Lesci\orcid{0000-0002-4607-2830}\inst{\ref{aff3},\ref{aff4}}
\and M.~Romanello\orcid{0000-0003-4563-4923}\inst{\ref{aff6},\ref{aff4}}
\and M.~Bolzonella\orcid{0000-0003-3278-4607}\inst{\ref{aff4}}
\and Z.~Sakr\orcid{0000-0002-4823-3757}\inst{\ref{aff7},\ref{aff8},\ref{aff9}}
\and B.~Altieri\orcid{0000-0003-3936-0284}\inst{\ref{aff10}}
\and S.~Andreon\orcid{0000-0002-2041-8784}\inst{\ref{aff11}}
\and C.~Baccigalupi\orcid{0000-0002-8211-1630}\inst{\ref{aff12},\ref{aff2},\ref{aff13},\ref{aff14}}
\and M.~Baldi\orcid{0000-0003-4145-1943}\inst{\ref{aff6},\ref{aff4},\ref{aff5}}
\and S.~Bardelli\orcid{0000-0002-8900-0298}\inst{\ref{aff4}}
\and P.~Battaglia\orcid{0000-0002-7337-5909}\inst{\ref{aff4}}
\and A.~Biviano\orcid{0000-0002-0857-0732}\inst{\ref{aff2},\ref{aff12}}
\and E.~Branchini\orcid{0000-0002-0808-6908}\inst{\ref{aff15},\ref{aff16},\ref{aff11}}
\and M.~Brescia\orcid{0000-0001-9506-5680}\inst{\ref{aff17},\ref{aff18}}
\and S.~Camera\orcid{0000-0003-3399-3574}\inst{\ref{aff19},\ref{aff20},\ref{aff21}}
\and V.~Capobianco\orcid{0000-0002-3309-7692}\inst{\ref{aff21}}
\and C.~Carbone\orcid{0000-0003-0125-3563}\inst{\ref{aff22}}
\and V.~F.~Cardone\inst{\ref{aff23},\ref{aff24}}
\and J.~Carretero\orcid{0000-0002-3130-0204}\inst{\ref{aff25},\ref{aff26}}
\and M.~Castellano\orcid{0000-0001-9875-8263}\inst{\ref{aff23}}
\and G.~Castignani\orcid{0000-0001-6831-0687}\inst{\ref{aff4}}
\and S.~Cavuoti\orcid{0000-0002-3787-4196}\inst{\ref{aff18},\ref{aff27}}
\and K.~C.~Chambers\orcid{0000-0001-6965-7789}\inst{\ref{aff28}}
\and A.~Cimatti\inst{\ref{aff29}}
\and C.~Colodro-Conde\inst{\ref{aff30}}
\and G.~Congedo\orcid{0000-0003-2508-0046}\inst{\ref{aff31}}
\and L.~Conversi\orcid{0000-0002-6710-8476}\inst{\ref{aff32},\ref{aff10}}
\and Y.~Copin\orcid{0000-0002-5317-7518}\inst{\ref{aff33}}
\and F.~Courbin\orcid{0000-0003-0758-6510}\inst{\ref{aff34},\ref{aff35},\ref{aff36}}
\and H.~M.~Courtois\orcid{0000-0003-0509-1776}\inst{\ref{aff37}}
\and H.~Degaudenzi\orcid{0000-0002-5887-6799}\inst{\ref{aff38}}
\and S.~de~la~Torre\inst{\ref{aff39}}
\and G.~De~Lucia\orcid{0000-0002-6220-9104}\inst{\ref{aff2}}
\and H.~Dole\orcid{0000-0002-9767-3839}\inst{\ref{aff40}}
\and F.~Dubath\orcid{0000-0002-6533-2810}\inst{\ref{aff38}}
\and X.~Dupac\inst{\ref{aff10}}
\and S.~Dusini\orcid{0000-0002-1128-0664}\inst{\ref{aff41}}
\and S.~Escoffier\orcid{0000-0002-2847-7498}\inst{\ref{aff42}}
\and M.~Farina\orcid{0000-0002-3089-7846}\inst{\ref{aff43}}
\and R.~Farinelli\inst{\ref{aff4}}
\and S.~Farrens\orcid{0000-0002-9594-9387}\inst{\ref{aff44}}
\and S.~Ferriol\inst{\ref{aff33}}
\and F.~Finelli\orcid{0000-0002-6694-3269}\inst{\ref{aff4},\ref{aff45}}
\and P.~Fosalba\orcid{0000-0002-1510-5214}\inst{\ref{aff46},\ref{aff47}}
\and S.~Fotopoulou\orcid{0000-0002-9686-254X}\inst{\ref{aff48}}
\and M.~Frailis\orcid{0000-0002-7400-2135}\inst{\ref{aff2}}
\and E.~Franceschi\orcid{0000-0002-0585-6591}\inst{\ref{aff4}}
\and M.~Fumana\orcid{0000-0001-6787-5950}\inst{\ref{aff22}}
\and S.~Galeotta\orcid{0000-0002-3748-5115}\inst{\ref{aff2}}
\and K.~George\orcid{0000-0002-1734-8455}\inst{\ref{aff49}}
\and B.~Gillis\orcid{0000-0002-4478-1270}\inst{\ref{aff31}}
\and C.~Giocoli\orcid{0000-0002-9590-7961}\inst{\ref{aff4},\ref{aff5}}
\and J.~Gracia-Carpio\orcid{0000-0003-4689-3134}\inst{\ref{aff50}}
\and A.~Grazian\orcid{0000-0002-5688-0663}\inst{\ref{aff51}}
\and F.~Grupp\inst{\ref{aff50},\ref{aff52}}
\and S.~V.~H.~Haugan\orcid{0000-0001-9648-7260}\inst{\ref{aff53}}
\and W.~Holmes\inst{\ref{aff54}}
\and F.~Hormuth\inst{\ref{aff55}}
\and A.~Hornstrup\orcid{0000-0002-3363-0936}\inst{\ref{aff56},\ref{aff57}}
\and K.~Jahnke\orcid{0000-0003-3804-2137}\inst{\ref{aff58}}
\and M.~Jhabvala\inst{\ref{aff59}}
\and B.~Joachimi\orcid{0000-0001-7494-1303}\inst{\ref{aff60}}
\and S.~Kermiche\orcid{0000-0002-0302-5735}\inst{\ref{aff42}}
\and A.~Kiessling\orcid{0000-0002-2590-1273}\inst{\ref{aff54}}
\and B.~Kubik\orcid{0009-0006-5823-4880}\inst{\ref{aff33}}
\and M.~K\"ummel\orcid{0000-0003-2791-2117}\inst{\ref{aff52}}
\and M.~Kunz\orcid{0000-0002-3052-7394}\inst{\ref{aff61}}
\and H.~Kurki-Suonio\orcid{0000-0002-4618-3063}\inst{\ref{aff62},\ref{aff63}}
\and A.~M.~C.~Le~Brun\orcid{0000-0002-0936-4594}\inst{\ref{aff64}}
\and S.~Ligori\orcid{0000-0003-4172-4606}\inst{\ref{aff21}}
\and P.~B.~Lilje\orcid{0000-0003-4324-7794}\inst{\ref{aff53}}
\and V.~Lindholm\orcid{0000-0003-2317-5471}\inst{\ref{aff62},\ref{aff63}}
\and I.~Lloro\orcid{0000-0001-5966-1434}\inst{\ref{aff65}}
\and G.~Mainetti\orcid{0000-0003-2384-2377}\inst{\ref{aff66}}
\and O.~Mansutti\orcid{0000-0001-5758-4658}\inst{\ref{aff2}}
\and O.~Marggraf\orcid{0000-0001-7242-3852}\inst{\ref{aff67}}
\and M.~Martinelli\orcid{0000-0002-6943-7732}\inst{\ref{aff23},\ref{aff24}}
\and N.~Martinet\orcid{0000-0003-2786-7790}\inst{\ref{aff39}}
\and R.~J.~Massey\orcid{0000-0002-6085-3780}\inst{\ref{aff68}}
\and S.~Maurogordato\inst{\ref{aff69}}
\and E.~Medinaceli\orcid{0000-0002-4040-7783}\inst{\ref{aff4}}
\and S.~Mei\orcid{0000-0002-2849-559X}\inst{\ref{aff70},\ref{aff71}}
\and M.~Meneghetti\orcid{0000-0003-1225-7084}\inst{\ref{aff4},\ref{aff5}}
\and E.~Merlin\orcid{0000-0001-6870-8900}\inst{\ref{aff23}}
\and G.~Meylan\inst{\ref{aff72}}
\and A.~Mora\orcid{0000-0002-1922-8529}\inst{\ref{aff73}}
\and M.~Moresco\orcid{0000-0002-7616-7136}\inst{\ref{aff3},\ref{aff4}}
\and L.~Moscardini\orcid{0000-0002-3473-6716}\inst{\ref{aff3},\ref{aff4},\ref{aff5}}
\and E.~Munari\orcid{0000-0002-1751-5946}\inst{\ref{aff2},\ref{aff12}}
\and C.~Neissner\orcid{0000-0001-8524-4968}\inst{\ref{aff74},\ref{aff26}}
\and S.-M.~Niemi\orcid{0009-0005-0247-0086}\inst{\ref{aff75}}
\and J.~W.~Nightingale\orcid{0000-0002-8987-7401}\inst{\ref{aff76}}
\and C.~Padilla\orcid{0000-0001-7951-0166}\inst{\ref{aff74}}
\and S.~Paltani\orcid{0000-0002-8108-9179}\inst{\ref{aff38}}
\and F.~Pasian\orcid{0000-0002-4869-3227}\inst{\ref{aff2}}
\and K.~Pedersen\inst{\ref{aff77}}
\and V.~Pettorino\orcid{0000-0002-4203-9320}\inst{\ref{aff75}}
\and S.~Pires\orcid{0000-0002-0249-2104}\inst{\ref{aff44}}
\and G.~Polenta\orcid{0000-0003-4067-9196}\inst{\ref{aff78}}
\and M.~Poncet\inst{\ref{aff79}}
\and L.~A.~Popa\inst{\ref{aff80}}
\and F.~Raison\orcid{0000-0002-7819-6918}\inst{\ref{aff50}}
\and J.~Rhodes\orcid{0000-0002-4485-8549}\inst{\ref{aff54}}
\and G.~Riccio\inst{\ref{aff18}}
\and E.~Romelli\orcid{0000-0003-3069-9222}\inst{\ref{aff2}}
\and M.~Roncarelli\orcid{0000-0001-9587-7822}\inst{\ref{aff4}}
\and R.~Saglia\orcid{0000-0003-0378-7032}\inst{\ref{aff52},\ref{aff50}}
\and D.~Sapone\orcid{0000-0001-7089-4503}\inst{\ref{aff81}}
\and B.~Sartoris\orcid{0000-0003-1337-5269}\inst{\ref{aff52},\ref{aff2}}
\and P.~Schneider\orcid{0000-0001-8561-2679}\inst{\ref{aff67}}
\and A.~Secroun\orcid{0000-0003-0505-3710}\inst{\ref{aff42}}
\and E.~Sihvola\orcid{0000-0003-1804-7715}\inst{\ref{aff82}}
\and P.~Simon\inst{\ref{aff67}}
\and C.~Sirignano\orcid{0000-0002-0995-7146}\inst{\ref{aff83},\ref{aff41}}
\and G.~Sirri\orcid{0000-0003-2626-2853}\inst{\ref{aff5}}
\and A.~Spurio~Mancini\orcid{0000-0001-5698-0990}\inst{\ref{aff84}}
\and L.~Stanco\orcid{0000-0002-9706-5104}\inst{\ref{aff41}}
\and P.~Tallada-Cresp\'{i}\orcid{0000-0002-1336-8328}\inst{\ref{aff25},\ref{aff26}}
\and A.~N.~Taylor\inst{\ref{aff31}}
\and I.~Tereno\orcid{0000-0002-4537-6218}\inst{\ref{aff85},\ref{aff86}}
\and N.~Tessore\orcid{0000-0002-9696-7931}\inst{\ref{aff87}}
\and S.~Toft\orcid{0000-0003-3631-7176}\inst{\ref{aff88},\ref{aff89}}
\and R.~Toledo-Moreo\orcid{0000-0002-2997-4859}\inst{\ref{aff90}}
\and F.~Torradeflot\orcid{0000-0003-1160-1517}\inst{\ref{aff26},\ref{aff25}}
\and I.~Tutusaus\orcid{0000-0002-3199-0399}\inst{\ref{aff47},\ref{aff46},\ref{aff8}}
\and J.~Valiviita\orcid{0000-0001-6225-3693}\inst{\ref{aff62},\ref{aff63}}
\and T.~Vassallo\orcid{0000-0001-6512-6358}\inst{\ref{aff2},\ref{aff49}}
\and G.~Verdoes~Kleijn\orcid{0000-0001-5803-2580}\inst{\ref{aff91}}
\and Y.~Wang\orcid{0000-0002-4749-2984}\inst{\ref{aff92}}
\and J.~Weller\orcid{0000-0002-8282-2010}\inst{\ref{aff52},\ref{aff50}}
\and G.~Zamorani\orcid{0000-0002-2318-301X}\inst{\ref{aff4}}
\and F.~M.~Zerbi\orcid{0000-0002-9996-973X}\inst{\ref{aff11}}
\and E.~Zucca\orcid{0000-0002-5845-8132}\inst{\ref{aff4}}
\and M.~Ballardini\orcid{0000-0003-4481-3559}\inst{\ref{aff93},\ref{aff94},\ref{aff4}}
\and C.~Benoist\inst{\ref{aff69}}
\and A.~Boucaud\orcid{0000-0001-7387-2633}\inst{\ref{aff70}}
\and E.~Bozzo\orcid{0000-0002-8201-1525}\inst{\ref{aff38}}
\and C.~Burigana\orcid{0000-0002-3005-5796}\inst{\ref{aff95},\ref{aff45}}
\and M.~Calabrese\orcid{0000-0002-2637-2422}\inst{\ref{aff96},\ref{aff22}}
\and T.~Castro\orcid{0000-0002-6292-3228}\inst{\ref{aff2},\ref{aff13},\ref{aff12},\ref{aff97}}
\and J.~A.~Escartin~Vigo\inst{\ref{aff50}}
\and L.~Gabarra\orcid{0000-0002-8486-8856}\inst{\ref{aff98}}
\and J.~Garc\'ia-Bellido\orcid{0000-0002-9370-8360}\inst{\ref{aff7}}
\and J.~Macias-Perez\orcid{0000-0002-5385-2763}\inst{\ref{aff99}}
\and R.~Maoli\orcid{0000-0002-6065-3025}\inst{\ref{aff100},\ref{aff23}}
\and J.~Mart\'{i}n-Fleitas\orcid{0000-0002-8594-569X}\inst{\ref{aff101}}
\and N.~Mauri\orcid{0000-0001-8196-1548}\inst{\ref{aff29},\ref{aff5}}
\and R.~B.~Metcalf\orcid{0000-0003-3167-2574}\inst{\ref{aff3},\ref{aff4}}
\and P.~Monaco\orcid{0000-0003-2083-7564}\inst{\ref{aff102},\ref{aff2},\ref{aff13},\ref{aff12}}
\and A.~A.~Nucita\inst{\ref{aff103},\ref{aff104},\ref{aff105}}
\and A.~Pezzotta\orcid{0000-0003-0726-2268}\inst{\ref{aff11}}
\and M.~P\"ontinen\orcid{0000-0001-5442-2530}\inst{\ref{aff62}}
\and I.~Risso\orcid{0000-0003-2525-7761}\inst{\ref{aff11},\ref{aff16}}
\and V.~Scottez\orcid{0009-0008-3864-940X}\inst{\ref{aff106},\ref{aff107}}
\and M.~Sereno\orcid{0000-0003-0302-0325}\inst{\ref{aff4},\ref{aff5}}
\and M.~Tenti\orcid{0000-0002-4254-5901}\inst{\ref{aff5}}
\and M.~Tucci\inst{\ref{aff38}}
\and M.~Viel\orcid{0000-0002-2642-5707}\inst{\ref{aff12},\ref{aff2},\ref{aff14},\ref{aff13},\ref{aff97}}
\and M.~Wiesmann\orcid{0009-0000-8199-5860}\inst{\ref{aff53}}
\and Y.~Akrami\orcid{0000-0002-2407-7956}\inst{\ref{aff7},\ref{aff108}}
\and I.~T.~Andika\orcid{0000-0001-6102-9526}\inst{\ref{aff52}}
\and G.~Angora\orcid{0000-0002-0316-6562}\inst{\ref{aff18},\ref{aff93}}
\and M.~Archidiacono\orcid{0000-0003-4952-9012}\inst{\ref{aff109},\ref{aff110}}
\and F.~Atrio-Barandela\orcid{0000-0002-2130-2513}\inst{\ref{aff111}}
\and E.~Aubourg\orcid{0000-0002-5592-023X}\inst{\ref{aff70},\ref{aff112}}
\and L.~Bazzanini\orcid{0000-0003-0727-0137}\inst{\ref{aff93},\ref{aff4}}
\and J.~Bel\inst{\ref{aff113}}
\and D.~Bertacca\orcid{0000-0002-2490-7139}\inst{\ref{aff83},\ref{aff51},\ref{aff41}}
\and M.~Bethermin\orcid{0000-0002-3915-2015}\inst{\ref{aff114}}
\and F.~Beutler\orcid{0000-0003-0467-5438}\inst{\ref{aff31}}
\and A.~Blanchard\orcid{0000-0001-8555-9003}\inst{\ref{aff8}}
\and L.~Blot\orcid{0000-0002-9622-7167}\inst{\ref{aff115},\ref{aff64}}
\and M.~Bonici\orcid{0000-0002-8430-126X}\inst{\ref{aff116},\ref{aff22}}
\and S.~Borgani\orcid{0000-0001-6151-6439}\inst{\ref{aff102},\ref{aff12},\ref{aff2},\ref{aff13},\ref{aff97}}
\and M.~L.~Brown\orcid{0000-0002-0370-8077}\inst{\ref{aff117}}
\and S.~Bruton\orcid{0000-0002-6503-5218}\inst{\ref{aff118}}
\and A.~Calabro\orcid{0000-0003-2536-1614}\inst{\ref{aff23}}
\and B.~Camacho~Quevedo\orcid{0000-0002-8789-4232}\inst{\ref{aff12},\ref{aff14},\ref{aff2}}
\and F.~Caro\orcid{0009-0003-1053-0507}\inst{\ref{aff23}}
\and C.~S.~Carvalho\inst{\ref{aff86}}
\and F.~Cogato\orcid{0000-0003-4632-6113}\inst{\ref{aff3},\ref{aff4}}
\and S.~Contarini\orcid{0000-0002-9843-723X}\inst{\ref{aff50}}
\and A.~R.~Cooray\orcid{0000-0002-3892-0190}\inst{\ref{aff119}}
\and M.~Costanzi\orcid{0000-0001-8158-1449}\inst{\ref{aff102},\ref{aff2},\ref{aff12}}
\and F.~De~Paolis\orcid{0000-0001-6460-7563}\inst{\ref{aff103},\ref{aff104},\ref{aff105}}
\and G.~Desprez\orcid{0000-0001-8325-1742}\inst{\ref{aff91}}
\and A.~D\'iaz-S\'anchez\orcid{0000-0003-0748-4768}\inst{\ref{aff120}}
\and S.~Di~Domizio\orcid{0000-0003-2863-5895}\inst{\ref{aff15},\ref{aff16}}
\and J.~M.~Diego\orcid{0000-0001-9065-3926}\inst{\ref{aff121}}
\and V.~Duret\orcid{0009-0009-0383-4960}\inst{\ref{aff42}}
\and M.~Y.~Elkhashab\orcid{0000-0001-9306-2603}\inst{\ref{aff102},\ref{aff2},\ref{aff13},\ref{aff12}}
\and A.~Enia\orcid{0000-0002-0200-2857}\inst{\ref{aff4}}
\and Y.~Fang\orcid{0000-0002-0334-6950}\inst{\ref{aff52}}
\and A.~Finoguenov\orcid{0000-0002-4606-5403}\inst{\ref{aff62}}
\and A.~Franco\orcid{0000-0002-4761-366X}\inst{\ref{aff104},\ref{aff103},\ref{aff105}}
\and K.~Ganga\orcid{0000-0001-8159-8208}\inst{\ref{aff70}}
\and T.~Gasparetto\orcid{0000-0002-7913-4866}\inst{\ref{aff23}}
\and E.~Gaztanaga\orcid{0000-0001-9632-0815}\inst{\ref{aff47},\ref{aff46},\ref{aff122}}
\and F.~Giacomini\orcid{0000-0002-3129-2814}\inst{\ref{aff5}}
\and F.~Gianotti\orcid{0000-0003-4666-119X}\inst{\ref{aff4}}
\and E.~J.~Gonzalez\orcid{0000-0002-0226-9893}\inst{\ref{aff123},\ref{aff124}}
\and G.~Gozaliasl\orcid{0000-0002-0236-919X}\inst{\ref{aff125},\ref{aff62}}
\and A.~Gruppuso\orcid{0000-0001-9272-5292}\inst{\ref{aff4},\ref{aff5}}
\and M.~Guidi\orcid{0000-0001-9408-1101}\inst{\ref{aff6},\ref{aff4}}
\and C.~M.~Gutierrez\orcid{0000-0001-7854-783X}\inst{\ref{aff30},\ref{aff126}}
\and A.~Hall\orcid{0000-0002-3139-8651}\inst{\ref{aff31}}
\and C.~Hern\'andez-Monteagudo\orcid{0000-0001-5471-9166}\inst{\ref{aff126},\ref{aff30}}
\and H.~Hildebrandt\orcid{0000-0002-9814-3338}\inst{\ref{aff127}}
\and J.~Hjorth\orcid{0000-0002-4571-2306}\inst{\ref{aff77}}
\and J.~J.~E.~Kajava\orcid{0000-0002-3010-8333}\inst{\ref{aff128},\ref{aff129},\ref{aff130}}
\and Y.~Kang\orcid{0009-0000-8588-7250}\inst{\ref{aff38}}
\and V.~Kansal\orcid{0000-0002-4008-6078}\inst{\ref{aff131},\ref{aff132}}
\and D.~Karagiannis\orcid{0000-0002-4927-0816}\inst{\ref{aff93},\ref{aff133}}
\and K.~Kiiveri\inst{\ref{aff82}}
\and J.~Kim\orcid{0000-0003-2776-2761}\inst{\ref{aff98}}
\and C.~C.~Kirkpatrick\inst{\ref{aff82}}
\and S.~Kruk\orcid{0000-0001-8010-8879}\inst{\ref{aff10}}
\and M.~Lattanzi\orcid{0000-0003-1059-2532}\inst{\ref{aff94}}
\and J.~Le~Graet\orcid{0000-0001-6523-7971}\inst{\ref{aff42}}
\and L.~Legrand\orcid{0000-0003-0610-5252}\inst{\ref{aff134},\ref{aff135}}
\and M.~Lembo\orcid{0000-0002-5271-5070}\inst{\ref{aff136}}
\and F.~Lepori\orcid{0009-0000-5061-7138}\inst{\ref{aff137}}
\and G.~Leroy\orcid{0009-0004-2523-4425}\inst{\ref{aff138},\ref{aff68}}
\and J.~Lesgourgues\orcid{0000-0001-7627-353X}\inst{\ref{aff139}}
\and T.~I.~Liaudat\orcid{0000-0002-9104-314X}\inst{\ref{aff112}}
\and S.~J.~Liu\orcid{0000-0001-7680-2139}\inst{\ref{aff43}}
\and A.~Loureiro\orcid{0000-0002-4371-0876}\inst{\ref{aff140},\ref{aff1}}
\and M.~Magliocchetti\orcid{0000-0001-9158-4838}\inst{\ref{aff43}}
\and A.~Manj\'on-Garc\'ia\orcid{0000-0002-7413-8825}\inst{\ref{aff120}}
\and F.~Mannucci\orcid{0000-0002-4803-2381}\inst{\ref{aff141}}
\and C.~J.~A.~P.~Martins\orcid{0000-0002-4886-9261}\inst{\ref{aff142},\ref{aff143}}
\and L.~Maurin\orcid{0000-0002-8406-0857}\inst{\ref{aff40}}
\and M.~Miluzio\inst{\ref{aff10},\ref{aff144}}
\and C.~Moretti\orcid{0000-0003-3314-8936}\inst{\ref{aff2},\ref{aff12},\ref{aff13}}
\and G.~Morgante\inst{\ref{aff4}}
\and S.~Nadathur\orcid{0000-0001-9070-3102}\inst{\ref{aff122}}
\and K.~Naidoo\orcid{0000-0002-9182-1802}\inst{\ref{aff122},\ref{aff58}}
\and A.~Navarro-Alsina\orcid{0000-0002-3173-2592}\inst{\ref{aff67}}
\and S.~Nesseris\orcid{0000-0002-0567-0324}\inst{\ref{aff7}}
\and D.~Paoletti\orcid{0000-0003-4761-6147}\inst{\ref{aff4},\ref{aff45}}
\and K.~Paterson\orcid{0000-0001-8340-3486}\inst{\ref{aff58}}
\and L.~Patrizii\inst{\ref{aff5}}
\and C.~Pattison\orcid{0000-0003-3272-2617}\inst{\ref{aff122}}
\and A.~Pisani\orcid{0000-0002-6146-4437}\inst{\ref{aff42}}
\and D.~Potter\orcid{0000-0002-0757-5195}\inst{\ref{aff145}}
\and G.~W.~Pratt\inst{\ref{aff44}}
\and S.~Quai\orcid{0000-0002-0449-8163}\inst{\ref{aff3},\ref{aff4}}
\and M.~Radovich\orcid{0000-0002-3585-866X}\inst{\ref{aff51}}
\and K.~Rojas\orcid{0000-0003-1391-6854}\inst{\ref{aff146}}
\and W.~Roster\orcid{0000-0002-9149-6528}\inst{\ref{aff50}}
\and S.~Sacquegna\orcid{0000-0002-8433-6630}\inst{\ref{aff147}}
\and M.~Sahl\'en\orcid{0000-0003-0973-4804}\inst{\ref{aff148}}
\and D.~B.~Sanders\orcid{0000-0002-1233-9998}\inst{\ref{aff28}}
\and E.~Sarpa\orcid{0000-0002-1256-655X}\inst{\ref{aff2}}
\and A.~Schneider\orcid{0000-0001-7055-8104}\inst{\ref{aff145}}
\and M.~Schultheis\inst{\ref{aff69}}
\and D.~Sciotti\orcid{0009-0008-4519-2620}\inst{\ref{aff23},\ref{aff24}}
\and E.~Sellentin\inst{\ref{aff149},\ref{aff150}}
\and L.~C.~Smith\orcid{0000-0002-3259-2771}\inst{\ref{aff151}}
\and J.~G.~Sorce\orcid{0000-0002-2307-2432}\inst{\ref{aff152},\ref{aff40}}
\and K.~Tanidis\orcid{0000-0001-9843-5130}\inst{\ref{aff153}}
\and F.~Tarsitano\orcid{0000-0002-5919-0238}\inst{\ref{aff154},\ref{aff38}}
\and G.~Testera\inst{\ref{aff16}}
\and R.~Teyssier\orcid{0000-0001-7689-0933}\inst{\ref{aff155}}
\and S.~Tosi\orcid{0000-0002-7275-9193}\inst{\ref{aff15},\ref{aff11},\ref{aff16}}
\and A.~Troja\orcid{0000-0003-0239-4595}\inst{\ref{aff2}}
\and C.~Valieri\inst{\ref{aff5}}
\and A.~Venhola\orcid{0000-0001-6071-4564}\inst{\ref{aff156}}
\and D.~Vergani\orcid{0000-0003-0898-2216}\inst{\ref{aff4}}
\and G.~Verza\orcid{0000-0002-1886-8348}\inst{\ref{aff157},\ref{aff158}}
\and S.~Vinciguerra\orcid{0009-0005-4018-3184}\inst{\ref{aff39}}
\and N.~A.~Walton\orcid{0000-0003-3983-8778}\inst{\ref{aff151}}
\and A.~H.~Wright\orcid{0000-0001-7363-7932}\inst{\ref{aff127}}}
										   
\institute{Astrophysics Group, Blackett Laboratory, Imperial College London, London SW7 2AZ, UK\label{aff1}
\and
INAF-Osservatorio Astronomico di Trieste, Via G. B. Tiepolo 11, 34143 Trieste, Italy\label{aff2}
\and
Dipartimento di Fisica e Astronomia "Augusto Righi" - Alma Mater Studiorum Universit\`a di Bologna, via Piero Gobetti 93/2, 40129 Bologna, Italy\label{aff3}
\and
INAF-Osservatorio di Astrofisica e Scienza dello Spazio di Bologna, Via Piero Gobetti 93/3, 40129 Bologna, Italy\label{aff4}
\and
INFN-Sezione di Bologna, Viale Berti Pichat 6/2, 40127 Bologna, Italy\label{aff5}
\and
Dipartimento di Fisica e Astronomia, Universit\`a di Bologna, Via Gobetti 93/2, 40129 Bologna, Italy\label{aff6}
\and
Instituto de F\'isica Te\'orica UAM-CSIC, Campus de Cantoblanco, 28049 Madrid, Spain\label{aff7}
\and
Institut de Recherche en Astrophysique et Plan\'etologie (IRAP), Universit\'e de Toulouse, CNRS, UPS, CNES, 14 Av. Edouard Belin, 31400 Toulouse, France\label{aff8}
\and
Universit\'e St Joseph; Faculty of Sciences, Beirut, Lebanon\label{aff9}
\and
ESAC/ESA, Camino Bajo del Castillo, s/n., Urb. Villafranca del Castillo, 28692 Villanueva de la Ca\~nada, Madrid, Spain\label{aff10}
\and
INAF-Osservatorio Astronomico di Brera, Via Brera 28, 20122 Milano, Italy\label{aff11}
\and
IFPU, Institute for Fundamental Physics of the Universe, via Beirut 2, 34151 Trieste, Italy\label{aff12}
\and
INFN, Sezione di Trieste, Via Valerio 2, 34127 Trieste TS, Italy\label{aff13}
\and
SISSA, International School for Advanced Studies, Via Bonomea 265, 34136 Trieste TS, Italy\label{aff14}
\and
Dipartimento di Fisica, Universit\`a di Genova, Via Dodecaneso 33, 16146, Genova, Italy\label{aff15}
\and
INFN-Sezione di Genova, Via Dodecaneso 33, 16146, Genova, Italy\label{aff16}
\and
Department of Physics "E. Pancini", University Federico II, Via Cinthia 6, 80126, Napoli, Italy\label{aff17}
\and
INAF-Osservatorio Astronomico di Capodimonte, Via Moiariello 16, 80131 Napoli, Italy\label{aff18}
\and
Dipartimento di Fisica, Universit\`a degli Studi di Torino, Via P. Giuria 1, 10125 Torino, Italy\label{aff19}
\and
INFN-Sezione di Torino, Via P. Giuria 1, 10125 Torino, Italy\label{aff20}
\and
INAF-Osservatorio Astrofisico di Torino, Via Osservatorio 20, 10025 Pino Torinese (TO), Italy\label{aff21}
\and
INAF-IASF Milano, Via Alfonso Corti 12, 20133 Milano, Italy\label{aff22}
\and
INAF-Osservatorio Astronomico di Roma, Via Frascati 33, 00078 Monteporzio Catone, Italy\label{aff23}
\and
INFN-Sezione di Roma, Piazzale Aldo Moro, 2 - c/o Dipartimento di Fisica, Edificio G. Marconi, 00185 Roma, Italy\label{aff24}
\and
Centro de Investigaciones Energ\'eticas, Medioambientales y Tecnol\'ogicas (CIEMAT), Avenida Complutense 40, 28040 Madrid, Spain\label{aff25}
\and
Port d'Informaci\'{o} Cient\'{i}fica, Campus UAB, C. Albareda s/n, 08193 Bellaterra (Barcelona), Spain\label{aff26}
\and
INFN section of Naples, Via Cinthia 6, 80126, Napoli, Italy\label{aff27}
\and
Institute for Astronomy, University of Hawaii, 2680 Woodlawn Drive, Honolulu, HI 96822, USA\label{aff28}
\and
Dipartimento di Fisica e Astronomia "Augusto Righi" - Alma Mater Studiorum Universit\`a di Bologna, Viale Berti Pichat 6/2, 40127 Bologna, Italy\label{aff29}
\and
Instituto de Astrof\'{\i}sica de Canarias, E-38205 La Laguna, Tenerife, Spain\label{aff30}
\and
Institute for Astronomy, University of Edinburgh, Royal Observatory, Blackford Hill, Edinburgh EH9 3HJ, UK\label{aff31}
\and
European Space Agency/ESRIN, Largo Galileo Galilei 1, 00044 Frascati, Roma, Italy\label{aff32}
\and
Universit\'e Claude Bernard Lyon 1, CNRS/IN2P3, IP2I Lyon, UMR 5822, Villeurbanne, F-69100, France\label{aff33}
\and
Institut de Ci\`{e}ncies del Cosmos (ICCUB), Universitat de Barcelona (IEEC-UB), Mart\'{i} i Franqu\`{e}s 1, 08028 Barcelona, Spain\label{aff34}
\and
Instituci\'o Catalana de Recerca i Estudis Avan\c{c}ats (ICREA), Passeig de Llu\'{\i}s Companys 23, 08010 Barcelona, Spain\label{aff35}
\and
Institut de Ciencies de l'Espai (IEEC-CSIC), Campus UAB, Carrer de Can Magrans, s/n Cerdanyola del Vall\'es, 08193 Barcelona, Spain\label{aff36}
\and
UCB Lyon 1, CNRS/IN2P3, IUF, IP2I Lyon, 4 rue Enrico Fermi, 69622 Villeurbanne, France\label{aff37}
\and
Department of Astronomy, University of Geneva, ch. d'Ecogia 16, 1290 Versoix, Switzerland\label{aff38}
\and
Aix-Marseille Universit\'e, CNRS, CNES, LAM, Marseille, France\label{aff39}
\and
Universit\'e Paris-Saclay, CNRS, Institut d'astrophysique spatiale, 91405, Orsay, France\label{aff40}
\and
INFN-Padova, Via Marzolo 8, 35131 Padova, Italy\label{aff41}
\and
Aix-Marseille Universit\'e, CNRS/IN2P3, CPPM, Marseille, France\label{aff42}
\and
INAF-Istituto di Astrofisica e Planetologia Spaziali, via del Fosso del Cavaliere, 100, 00100 Roma, Italy\label{aff43}
\and
Universit\'e Paris-Saclay, Universit\'e Paris Cit\'e, CEA, CNRS, AIM, 91191, Gif-sur-Yvette, France\label{aff44}
\and
INFN-Bologna, Via Irnerio 46, 40126 Bologna, Italy\label{aff45}
\and
Institut d'Estudis Espacials de Catalunya (IEEC),  Edifici RDIT, Campus UPC, 08860 Castelldefels, Barcelona, Spain\label{aff46}
\and
Institute of Space Sciences (ICE, CSIC), Campus UAB, Carrer de Can Magrans, s/n, 08193 Barcelona, Spain\label{aff47}
\and
School of Physics, HH Wills Physics Laboratory, University of Bristol, Tyndall Avenue, Bristol, BS8 1TL, UK\label{aff48}
\and
University Observatory, LMU Faculty of Physics, Scheinerstr.~1, 81679 Munich, Germany\label{aff49}
\and
Max Planck Institute for Extraterrestrial Physics, Giessenbachstr. 1, 85748 Garching, Germany\label{aff50}
\and
INAF-Osservatorio Astronomico di Padova, Via dell'Osservatorio 5, 35122 Padova, Italy\label{aff51}
\and
Universit\"ats-Sternwarte M\"unchen, Fakult\"at f\"ur Physik, Ludwig-Maximilians-Universit\"at M\"unchen, Scheinerstr.~1, 81679 M\"unchen, Germany\label{aff52}
\and
Institute of Theoretical Astrophysics, University of Oslo, P.O. Box 1029 Blindern, 0315 Oslo, Norway\label{aff53}
\and
Jet Propulsion Laboratory, California Institute of Technology, 4800 Oak Grove Drive, Pasadena, CA, 91109, USA\label{aff54}
\and
Felix Hormuth Engineering, Goethestr. 17, 69181 Leimen, Germany\label{aff55}
\and
Technical University of Denmark, Elektrovej 327, 2800 Kgs. Lyngby, Denmark\label{aff56}
\and
Cosmic Dawn Center (DAWN), Denmark\label{aff57}
\and
Max-Planck-Institut f\"ur Astronomie, K\"onigstuhl 17, 69117 Heidelberg, Germany\label{aff58}
\and
NASA Goddard Space Flight Center, Greenbelt, MD 20771, USA\label{aff59}
\and
Department of Physics and Astronomy, University College London, Gower Street, London WC1E 6BT, UK\label{aff60}
\and
Universit\'e de Gen\`eve, D\'epartement de Physique Th\'eorique and Centre for Astroparticle Physics, 24 quai Ernest-Ansermet, CH-1211 Gen\`eve 4, Switzerland\label{aff61}
\and
Department of Physics, P.O. Box 64, University of Helsinki, 00014 Helsinki, Finland\label{aff62}
\and
Helsinki Institute of Physics, Gustaf H{\"a}llstr{\"o}min katu 2, University of Helsinki, 00014 Helsinki, Finland\label{aff63}
\and
Laboratoire d'etude de l'Univers et des phenomenes eXtremes, Observatoire de Paris, Universit\'e PSL, Sorbonne Universit\'e, CNRS, 92190 Meudon, France\label{aff64}
\and
SKAO, Jodrell Bank, Lower Withington, Macclesfield SK11 9FT, UK\label{aff65}
\and
Centre de Calcul de l'IN2P3/CNRS, 21 avenue Pierre de Coubertin 69627 Villeurbanne Cedex, France\label{aff66}
\and
Universit\"at Bonn, Argelander-Institut f\"ur Astronomie, Auf dem H\"ugel 71, 53121 Bonn, Germany\label{aff67}
\and
Department of Physics, Institute for Computational Cosmology, Durham University, South Road, Durham, DH1 3LE, UK\label{aff68}
\and
Universit\'e C\^{o}te d'Azur, Observatoire de la C\^{o}te d'Azur, CNRS, Laboratoire Lagrange, Bd de l'Observatoire, CS 34229, 06304 Nice cedex 4, France\label{aff69}
\and
Universit\'e Paris Cit\'e, CNRS, Astroparticule et Cosmologie, 75013 Paris, France\label{aff70}
\and
CNRS-UCB International Research Laboratory, Centre Pierre Bin\'etruy, IRL2007, CPB-IN2P3, Berkeley, USA\label{aff71}
\and
Institute of Physics, Laboratory of Astrophysics, Ecole Polytechnique F\'ed\'erale de Lausanne (EPFL), Observatoire de Sauverny, 1290 Versoix, Switzerland\label{aff72}
\and
Telespazio UK S.L. for European Space Agency (ESA), Camino bajo del Castillo, s/n, Urbanizacion Villafranca del Castillo, Villanueva de la Ca\~nada, 28692 Madrid, Spain\label{aff73}
\and
Institut de F\'{i}sica d'Altes Energies (IFAE), The Barcelona Institute of Science and Technology, Campus UAB, 08193 Bellaterra (Barcelona), Spain\label{aff74}
\and
European Space Agency/ESTEC, Keplerlaan 1, 2201 AZ Noordwijk, The Netherlands\label{aff75}
\and
School of Mathematics, Statistics and Physics, Newcastle University, Herschel Building, Newcastle-upon-Tyne, NE1 7RU, UK\label{aff76}
\and
DARK, Niels Bohr Institute, University of Copenhagen, Jagtvej 155, 2200 Copenhagen, Denmark\label{aff77}
\and
Space Science Data Center, Italian Space Agency, via del Politecnico snc, 00133 Roma, Italy\label{aff78}
\and
Centre National d'Etudes Spatiales -- Centre spatial de Toulouse, 18 avenue Edouard Belin, 31401 Toulouse Cedex 9, France\label{aff79}
\and
Institute of Space Science, Str. Atomistilor, nr. 409 M\u{a}gurele, Ilfov, 077125, Romania\label{aff80}
\and
Departamento de F\'isica, FCFM, Universidad de Chile, Blanco Encalada 2008, Santiago, Chile\label{aff81}
\and
Department of Physics and Helsinki Institute of Physics, Gustaf H\"allstr\"omin katu 2, University of Helsinki, 00014 Helsinki, Finland\label{aff82}
\and
Dipartimento di Fisica e Astronomia "G. Galilei", Universit\`a di Padova, Via Marzolo 8, 35131 Padova, Italy\label{aff83}
\and
Department of Physics, Royal Holloway, University of London, Surrey TW20 0EX, UK\label{aff84}
\and
Departamento de F\'isica, Faculdade de Ci\^encias, Universidade de Lisboa, Edif\'icio C8, Campo Grande, PT1749-016 Lisboa, Portugal\label{aff85}
\and
Instituto de Astrof\'isica e Ci\^encias do Espa\c{c}o, Faculdade de Ci\^encias, Universidade de Lisboa, Tapada da Ajuda, 1349-018 Lisboa, Portugal\label{aff86}
\and
Mullard Space Science Laboratory, University College London, Holmbury St Mary, Dorking, Surrey RH5 6NT, UK\label{aff87}
\and
Cosmic Dawn Center (DAWN)\label{aff88}
\and
Niels Bohr Institute, University of Copenhagen, Jagtvej 128, 2200 Copenhagen, Denmark\label{aff89}
\and
Universidad Polit\'ecnica de Cartagena, Departamento de Electr\'onica y Tecnolog\'ia de Computadoras,  Plaza del Hospital 1, 30202 Cartagena, Spain\label{aff90}
\and
Kapteyn Astronomical Institute, University of Groningen, PO Box 800, 9700 AV Groningen, The Netherlands\label{aff91}
\and
Caltech/IPAC, 1200 E. California Blvd., Pasadena, CA 91125, USA\label{aff92}
\and
Dipartimento di Fisica e Scienze della Terra, Universit\`a degli Studi di Ferrara, Via Giuseppe Saragat 1, 44122 Ferrara, Italy\label{aff93}
\and
Istituto Nazionale di Fisica Nucleare, Sezione di Ferrara, Via Giuseppe Saragat 1, 44122 Ferrara, Italy\label{aff94}
\and
INAF, Istituto di Radioastronomia, Via Piero Gobetti 101, 40129 Bologna, Italy\label{aff95}
\and
Astronomical Observatory of the Autonomous Region of the Aosta Valley (OAVdA), Loc. Lignan 39, I-11020, Nus (Aosta Valley), Italy\label{aff96}
\and
ICSC - Centro Nazionale di Ricerca in High Performance Computing, Big Data e Quantum Computing, Via Magnanelli 2, Bologna, Italy\label{aff97}
\and
Department of Physics, Oxford University, Keble Road, Oxford OX1 3RH, UK\label{aff98}
\and
Univ. Grenoble Alpes, CNRS, Grenoble INP, LPSC-IN2P3, 53, Avenue des Martyrs, 38000, Grenoble, France\label{aff99}
\and
Dipartimento di Fisica, Sapienza Universit\`a di Roma, Piazzale Aldo Moro 2, 00185 Roma, Italy\label{aff100}
\and
Aurora Technology for European Space Agency (ESA), Camino bajo del Castillo, s/n, Urbanizacion Villafranca del Castillo, Villanueva de la Ca\~nada, 28692 Madrid, Spain\label{aff101}
\and
Dipartimento di Fisica - Sezione di Astronomia, Universit\`a di Trieste, Via Tiepolo 11, 34131 Trieste, Italy\label{aff102}
\and
Department of Mathematics and Physics E. De Giorgi, University of Salento, Via per Arnesano, CP-I93, 73100, Lecce, Italy\label{aff103}
\and
INFN, Sezione di Lecce, Via per Arnesano, CP-193, 73100, Lecce, Italy\label{aff104}
\and
INAF-Sezione di Lecce, c/o Dipartimento Matematica e Fisica, Via per Arnesano, 73100, Lecce, Italy\label{aff105}
\and
Institut d'Astrophysique de Paris, 98bis Boulevard Arago, 75014, Paris, France\label{aff106}
\and
ICL, Junia, Universit\'e Catholique de Lille, LITL, 59000 Lille, France\label{aff107}
\and
CERCA/ISO, Department of Physics, Case Western Reserve University, 10900 Euclid Avenue, Cleveland, OH 44106, USA\label{aff108}
\and
Dipartimento di Fisica "Aldo Pontremoli", Universit\`a degli Studi di Milano, Via Celoria 16, 20133 Milano, Italy\label{aff109}
\and
INFN-Sezione di Milano, Via Celoria 16, 20133 Milano, Italy\label{aff110}
\and
Departamento de F{\'\i}sica Fundamental. Universidad de Salamanca. Plaza de la Merced s/n. 37008 Salamanca, Spain\label{aff111}
\and
IRFU, CEA, Universit\'e Paris-Saclay 91191 Gif-sur-Yvette Cedex, France\label{aff112}
\and
Aix-Marseille Universit\'e, Universit\'e de Toulon, CNRS, CPT, Marseille, France\label{aff113}
\and
Universit\'e de Strasbourg, CNRS, Observatoire astronomique de Strasbourg, UMR 7550, 67000 Strasbourg, France\label{aff114}
\and
Center for Data-Driven Discovery, Kavli IPMU (WPI), UTIAS, The University of Tokyo, Kashiwa, Chiba 277-8583, Japan\label{aff115}
\and
Waterloo Centre for Astrophysics, University of Waterloo, Waterloo, Ontario N2L 3G1, Canada\label{aff116}
\and
Jodrell Bank Centre for Astrophysics, Department of Physics and Astronomy, University of Manchester, Oxford Road, Manchester M13 9PL, UK\label{aff117}
\and
California Institute of Technology, 1200 E California Blvd, Pasadena, CA 91125, USA\label{aff118}
\and
Department of Physics \& Astronomy, University of California Irvine, Irvine CA 92697, USA\label{aff119}
\and
Departamento F\'isica Aplicada, Universidad Polit\'ecnica de Cartagena, Campus Muralla del Mar, 30202 Cartagena, Murcia, Spain\label{aff120}
\and
Instituto de F\'isica de Cantabria, Edificio Juan Jord\'a, Avenida de los Castros, 39005 Santander, Spain\label{aff121}
\and
Institute of Cosmology and Gravitation, University of Portsmouth, Portsmouth PO1 3FX, UK\label{aff122}
\and
Departament de F\'{\i}sica, Universitat Aut\`onoma de Barcelona, 08193 Bellaterra (Barcelona), Spain\label{aff123}
\and
Instituto de Astronomia Teorica y Experimental (IATE-CONICET), Laprida 854, X5000BGR, C\'ordoba, Argentina\label{aff124}
\and
Department of Computer Science, Aalto University, PO Box 15400, Espoo, FI-00 076, Finland\label{aff125}
\and
Universidad de La Laguna, Dpto. Astrof\'\i sica, E-38206 La Laguna, Tenerife, Spain\label{aff126}
\and
Ruhr University Bochum, Faculty of Physics and Astronomy, Astronomical Institute (AIRUB), German Centre for Cosmological Lensing (GCCL), 44780 Bochum, Germany\label{aff127}
\and
Department of Physics and Astronomy, Vesilinnantie 5, University of Turku, 20014 Turku, Finland\label{aff128}
\and
Finnish Centre for Astronomy with ESO (FINCA), Quantum, Vesilinnantie 5, University of Turku, 20014 Turku, Finland\label{aff129}
\and
Serco for European Space Agency (ESA), Camino bajo del Castillo, s/n, Urbanizacion Villafranca del Castillo, Villanueva de la Ca\~nada, 28692 Madrid, Spain\label{aff130}
\and
ARC Centre of Excellence for Dark Matter Particle Physics, Melbourne, Australia\label{aff131}
\and
Centre for Astrophysics \& Supercomputing, Swinburne University of Technology,  Hawthorn, Victoria 3122, Australia\label{aff132}
\and
Department of Physics and Astronomy, University of the Western Cape, Bellville, Cape Town, 7535, South Africa\label{aff133}
\and
DAMTP, Centre for Mathematical Sciences, Wilberforce Road, Cambridge CB3 0WA, UK\label{aff134}
\and
Kavli Institute for Cosmology Cambridge, Madingley Road, Cambridge, CB3 0HA, UK\label{aff135}
\and
Institut d'Astrophysique de Paris, UMR 7095, CNRS, and Sorbonne Universit\'e, 98 bis boulevard Arago, 75014 Paris, France\label{aff136}
\and
Departement of Theoretical Physics, University of Geneva, Switzerland\label{aff137}
\and
Department of Physics, Centre for Extragalactic Astronomy, Durham University, South Road, Durham, DH1 3LE, UK\label{aff138}
\and
Institute for Theoretical Particle Physics and Cosmology (TTK), RWTH Aachen University, 52056 Aachen, Germany\label{aff139}
\and
Oskar Klein Centre for Cosmoparticle Physics, Department of Physics, Stockholm University, Stockholm, SE-106 91, Sweden\label{aff140}
\and
INAF-Osservatorio Astrofisico di Arcetri, Largo E. Fermi 5, 50125, Firenze, Italy\label{aff141}
\and
Centro de Astrof\'{\i}sica da Universidade do Porto, Rua das Estrelas, 4150-762 Porto, Portugal\label{aff142}
\and
Instituto de Astrof\'isica e Ci\^encias do Espa\c{c}o, Universidade do Porto, CAUP, Rua das Estrelas, PT4150-762 Porto, Portugal\label{aff143}
\and
HE Space for European Space Agency (ESA), Camino bajo del Castillo, s/n, Urbanizacion Villafranca del Castillo, Villanueva de la Ca\~nada, 28692 Madrid, Spain\label{aff144}
\and
Department of Astrophysics, University of Zurich, Winterthurerstrasse 190, 8057 Zurich, Switzerland\label{aff145}
\and
University of Applied Sciences and Arts of Northwestern Switzerland, School of Computer Science, 5210 Windisch, Switzerland\label{aff146}
\and
INAF - Osservatorio Astronomico d'Abruzzo, Via Maggini, 64100, Teramo, Italy\label{aff147}
\and
Theoretical astrophysics, Department of Physics and Astronomy, Uppsala University, Box 516, 751 37 Uppsala, Sweden\label{aff148}
\and
Mathematical Institute, University of Leiden, Einsteinweg 55, 2333 CA Leiden, The Netherlands\label{aff149}
\and
Leiden Observatory, Leiden University, Einsteinweg 55, 2333 CC Leiden, The Netherlands\label{aff150}
\and
Institute of Astronomy, University of Cambridge, Madingley Road, Cambridge CB3 0HA, UK\label{aff151}
\and
Univ. Lille, CNRS, Centrale Lille, UMR 9189 CRIStAL, 59000 Lille, France\label{aff152}
\and
Center for Astrophysics and Cosmology, University of Nova Gorica, Nova Gorica, Slovenia\label{aff153}
\and
Institute for Particle Physics and Astrophysics, Dept. of Physics, ETH Zurich, Wolfgang-Pauli-Strasse 27, 8093 Zurich, Switzerland\label{aff154}
\and
Department of Astrophysical Sciences, Peyton Hall, Princeton University, Princeton, NJ 08544, USA\label{aff155}
\and
Space physics and astronomy research unit, University of Oulu, Pentti Kaiteran katu 1, FI-90014 Oulu, Finland\label{aff156}
\and
International Centre for Theoretical Physics (ICTP), Strada Costiera 11, 34151 Trieste, Italy\label{aff157}
\and
Center for Computational Astrophysics, Flatiron Institute, 162 5th Avenue, 10010, New York, NY, USA\label{aff158}}    

\authorrunning{Euclid Collaboration: E. Tsaprazi et al.}

   \date{Received ???; accepted ???}
 
  \abstract
   {The \textit{Euclid} satellite will deliver a catalogue of optically selected galaxy clusters spanning from around $2000$ deg$^2$ in Data Release (DR) 1 to around $14\,000$ deg$^2$ in DR3. In this work, we assess the validity of cluster clustering (CC) models for template-fitting, which complements the full-shape methodology by providing cosmological information from the anisotropy of the redshift-space two-point correlation function (2PCF). Both methods will be used to analyse the cluster 2PCF multipoles with \textit{Euclid}. We examined the multipoles of the two-point redshift-space clustering of galaxy clusters simulated with the semi-analytic \texttt{PINOCCHIO} code using third-order Lagrangian perturbation theory, assuming a \textit{Euclid} DR1-like footprint of 500 deg$^2$ in the northern hemisphere and 1400 deg$^2$ in the southern hemisphere. We estimated the first three even multipoles of the 2PCF and associated covariance matrix from 1000 DR1-like synthetic catalogues. We studied the impact of modelling the relevant non-linearities, halo bias, and photometric redshift uncertainties on the 2PCF. We applied three clustering models to the mock catalogues at $0<z<2$ and with a virial mass of $M_{\rm vir}>10^{14}\;h^{-1}\,M_\odot$ under realistic and optimistic photometric redshift uncertainty scenarios. We formulated a set of permissive and conservative criteria that ought to be fulfilled by the multipole cut-off scales and validated them against 100 mock catalogues via an inference of the growth rate multiplied by the matter power spectrum normalisation parameter, $f\sigma_8$. We tested the dispersion, Scoccimarro, and Taruya--Nishimichi--Saito models. We find that the simplest of the three (i.e. the dispersion model) yields unbiased inferences on $f\sigma_8$ from CC down to $10$ $h^{-1}$ Mpc in a DR1-like setting. All clustering models provide very similar goodness-of-fit metrics in the presence of DR1-like cluster redshift uncertainties.}


   \maketitle

\section{Introduction}
Galaxy clusters are the most massive structures in the Universe to have undergone virialisation. Their clustering offers a sensitive probe of our cosmological model \citep{2009ApJ...692..265E,2010MNRAS.401.2477H,2012ApJ...749...81H}. Although clusters are less populous than galaxies, their strong clustering signal and high masses offer an independent handle on the large-scale structure. Therefore, the cluster clustering (CC) signal is in principle stronger than that of galaxies, provided that dense cluster samples are available \citep{2021ApJ...920...13M}. Furthermore, the impact of small-scale velocity dispersion on the CC signal is relatively small, because cluster centroids exhibit a low internal velocity dispersion and are less affected by small-scale random motions than galaxies \citep{2012A&A...547A.100V}. Therefore, CC can be viewed as a probe complementary to galaxy clustering, particularly with the advent of the high signal-to-noise ratio (S/N) Stage-IV surveys \citep{LSST, Euclid,2016MNRAS.459.1764S,2025A&A...697A...1E}.\\
\indent Unbiased cosmological constraints demand CC models that achieve sufficient accuracy across a wide range of scales and redshifts \citep[for example][]{1999MNRAS.305..866B,2013MNRAS.434..684M,2015MNRAS.449.4147S,2021ApJ...920...13M,2022A&A...665A.100L,2024A&A...682A.148F,2024A&A...682A..72R}. Since observations are obtained in redshift space, the primary modelling challenge in CC analyses is having an accurate model for redshift-space clustering anisotropies. The Kaiser model \citep{1987MNRAS.227....1K, 2001Natur.410..169P} provides the linear theory predictions in the plane-parallel limit \citep{2012A&A...547A.100V}, but extensions to this model have been proposed to account for non-linear effects at smaller scales (see e.g. \citealt{pezzotta} and \citealt{2021ApJ...920...13M} for a summary of these models). 

In this paper, we aim to test the performance of the redshift-space CC models for a \Euclid-like survey \citep{2025A&A...697A...1E} to inform the choice of CC model to be implemented in the Euclid Cosmology Likelihood for Observables \citep[CLOE;][]{cloe1,cloe2} for DR1. We also aim to formulate and calibrate a statistical framework for the determination of the empirical cut-off scale of the two-point correlation function of cluster clustering. Given the significant uncertainties in a DR1-like setting, we adopted a permissive and a conservative set of statistical tests, the latter to be considered for future DRs. The permissive criteria aim for acceptable goodness of fit and unbiasedness of the inferred $f\sigma_8$ posteriors given a cut-off scale, which is common for all multipoles at a given redshift. The conservative criteria further require that the estimated uncertainties accurately reflect the true scatter in the measurements and provide correct coverage, with a cut-off scale that is allowed to vary with multipole order. We based our main conclusions on the former set of criteria. In this analysis, we calibrated our statistical requirements against simulations and we plan to repeat this analysis on real data when they become available.

We used the semi-analytic \code{PINOCCHIO} catalogues \citep{pinocchio,2017MNRAS.465.4658M}, which we used to extract 1000 light cones and one random catalogue, and to measure the corresponding 2PCF multipoles, specifically the monopole, quadrupole, and hexadecapole. We compared these against predictions from clustering models to determine the simplest model that achieves the required accuracy for the \textit{Euclid} pipeline. We did not consider odd multipoles because in the auto-correlation of a single tracer in the plane-parallel approximation, they ended up vanishing as a result of symmetry. We assume cluster photometric redshift uncertainties of $\sigma_\mathrm{z}=0.005(1+z)$ in the optimistic scenario and $\sigma_\mathrm{z}=0.01(1+z)$ in the realistic one, which are among the values indicated in \citet{Q1-SP050}. These values are determined by considering the photometric redshift uncertainties of stage-III surveys \citep{2025PhRvD.112h3535A}. For simplicity, we modelled cluster redshift uncertainties as a Gaussian dispersion. Although real photometric cluster redshift errors can exhibit non-Gaussian tails or catastrophic outliers, the dominant effect on the clustering signal is the line-of-sight damping produced by the overall dispersion. As long as the effective width of the error distribution is similar, the impact on large-scale clustering should be comparable. A more realistic redshift error distribution would primarily alter the precise form of the damping kernel, but it is unlikely to qualitatively affect our conclusions on the scales considered. We quantified the accuracy of the \citet{1994MNRAS.267.1020P}, \citet{2004PhRvD..70h3007S}, and \citet{2010PhRvD..82f3522T} models at separations 10--150 \Mpch with an additional view to investigate what modelling improvements are required for upcoming DRs. We chose this maximum scale to ensure that the baryon acoustic oscillation (BAO) features would be fully visible for large photometric redshift uncertainties. Scales larger than the BAO scale do not add any substantial cosmological information \citep{2024A&A...682A.148F,2025arXiv251013509F}.

\citet{2024A&A...683A.253E} and \citet{2025A&A...697A.140F} showed that the semi-analytic \code{PINOCCHIO} catalogues accurately reproduce the halo 2PCF covariance of the monopole obtained from $N$-body simulations in redshift space. More recently, \citet{2025arXiv251013509F} analysed the impact of systematic effects on the monopole of CC using \code{PINOCCHIO} and demonstrated that neglecting non-linear contributions can significantly bias the inferred $\Omega_\mathrm{m}$–$\sigma_8$ posterior. To obtain unbiased cosmological posteriors, the authors of that study mitigated the modelling mis-specifications by calibrating the monopole to the model prediction down to $20$ \Mpch in mocks covering $10\,313$\,deg$^2$ (between DR2 and DR3 footprint). In this work, we address the complementary issue: pinpointing the minimum cut-off scale that would ensure an unbiased inference when using the multipoles of the 2PCF. Our results directly inform the implementation of the CC model in \code{CLOE} and the associated analysis choices for cosmological inference. Since \code{PINOCCHIO} uses Lagrangian perturbation theory (LPT), we expect it to produce predictions that differ from those of an $N$-body simulation, potentially in a scale-dependent fashion. These predictions will also be highly uncertain for halo bias. However, for the purposes of this study, which is meant to provide a statistical framework for the determination of the cut-off scale in real data, we chose \code{PINOCCHIO}, whose covariance is well-understood \citep{2025A&A...697A.140F}. The multiple realisations offered by \code{PINOCCHIO} further allow us to monitor the average behaviour of our inferences. For the real data application in the future, we plan to repeat this analysis using the tests described in Sect. \ref{sec:criteria} as a guide.

We focussed on inferring $f\sigma_8$ as a diagnostic variable for validation because our analysis centres on the quadrupole. \citet{karcher} performed a related cut-off analysis for galaxy-clustering multipoles, assuming a common scale cut for all multipoles in the limit of no redshift uncertainties or angular–radial systematic effects. Here, in our set of conservative statistical tests, we allow for various scale cuts for different multipoles, cluster redshift uncertainties in the presence of angular systematic effects; however, we restrict our analysis to simpler clustering models, which we find to be adequate for photometric clustering. Scanning the parameter space of plausible cut-off scales that vary with redshift, redshift uncertainty, and multipole order across many mock catalogues is computationally expensive. To reduce this cost, we adopt a template-fitting strategy. Unlike full-shape analyses like the one for the monopole of the CC 2PCF by \citet{2025arXiv251013509F}, template fitting keeps the matter power spectrum fixed to a fiducial cosmology and varies only the shape-preserving parameters that rescale or anisotropically distort it. In our parameterisation, we use the growth rate times the matter power spectrum normalisation parameter, $f\sigma_8$, the linear halo bias times the matter power spectrum normalisation parameter, $b\sigma_8$, a velocity dispersion scale, $\sigma_\mathrm{v}$, and the parallel and perpendicular BAO parameters, $\alpha_\parallel$, and $\alpha_\perp$, respectively. We note that alternative parameterisations, such as the one proposed by \citet[][Eq. 4.10]{2025JCAP...09..008A}, could also be explored in future analyses. However, given the relatively large uncertainties expected for cluster clustering in DR1, we did not anticipate significant differences arising from the choice of parameterisation for the amplitude of redshift-space distortions.

Template fitting is fast, robust to small deviations from the fiducial cosmology, which avoids repeated evaluations of the matter power spectrum, halo mass function, and halo bias. A limitation of this approach is that cosmological parameters such as $\Omega_\mathrm{m}$ and $\sigma_8$ are not varied. Here, we keep the shape of the matter power spectrum, the halo bias, and the distance–redshift relation fixed to a fiducial cosmology, while $\Omega_\mathrm{m}$ and $\sigma_8$ are in principle constrained via the BAO parameters. As a result, any cosmology dependence of the effective halo bias must be absorbed into the nuisance parameter, $b\sigma_8$. For clusters, however, the degeneracy between effective halo bias and $\sigma_8$ is substantially reduced because the selection function, constrained by the observable–mass relation, its scatter, weak lensing calibration, and number counts, is far better determined than in galaxy-clustering analyses (see \citealt{2022A&A...665A.100L} and references therein).

To emulate a setting of well-understood systematic effects, we limited our analysis to a regime where the mass-redshift completeness is almost uniform. To this end, we impose a mass cut at $M_{\rm vir}>10^{14}\,h^{-1}M_\odot$, where sample uncertainties are better characterised. Second, we fixed $b\sigma_8$ and studied the sensitivity of the resulting cut-off scales to the assumed halo-bias prescription, using the \citetalias{tinker} and \citetalias{2024AA...691A..62E} prescriptions. A prior on $b\sigma_8$, as is expected in real observations, weakens the predictive power of statistical tests on the cut-off scale. In our analysis, we chose to fix $b\sigma_8$ because (i) a strong prior on halo bias is expected; (ii) \code{PINOCCHIO} provides only an approximate mapping between dark matter and halos; (iii) the determination of the cut-off scale becomes more robust to the statistical test configuration in less noisy datasets with more conservative error bars; (iv) our goal is a model test rather than a forecast; and (v) the halo bias is a nuisance parameter that only modulates the amplitude of the 2PCF.

Determining which redshift-space CC model performs best in a given setting is crucial, as it impacts all studies that derive cosmological constraints from cluster clustering statistics \citep{2016MNRAS.458.1909V,2018A&A...620A...1M,2021ApJ...919..144M,2023A&A...674A..80L,2025A&A...693A.195R}. This is especially true on quasi- and non-linear scales ($r < 20\,h^{-1}\mathrm{Mpc}$), where different redshift-space prescriptions begin to diverge and where a large fraction of the cosmological information resides. Pinning down the regime of validity of each model and, where possible, extending it deeper into the non-linear regime; therefore, it is essential both to avoid biased constraints and to fully exploit the information content of cluster clustering. Further, it is necessary to understand which regimes each model performs best in, so that we can accurately quantify potential systematic effects in real data applications. Finally, even though state-of-the-art redshift-space clustering models have achieved a high degree of accuracy for galaxy surveys \citep{karcher}, such a level of sophistication may not be required for photometric CC, where statistical uncertainties are higher. Given the computational cost of the most advanced models, it is therefore important to determine which simpler formulations are sufficiently accurate for cluster catalogues under DR1-like statistical uncertainties.

The Sloan Digital Sky Survey has provided constraints on the growth of structure from CC at intermediate redshifts, both at the level of spectroscopic clustering \citep{2021ApJ...920...13M} and with photometrically selected samples \citep{2015MNRAS.449.4147S,2024A&A...682A.148F}. These studies have highlighted the need for more extensive cluster datasets with improved redshift characterisation. \textit{Euclid} will detect around $60\,000$ clusters over an area of $14\,000$ deg$^2$ \citep{Euclid,2025A&A...697A...1E}, which will become available in the Euclid Catalogue of Galaxy Clusters (ECGC). The sample is expected to have a purity of $>80\%$ down to $10^{14}$ $h^{-1}M_\odot$  and up to redshift $z=2$ \citep{adam19}. Those properties will depend on the quality of the photometric redshift estimates. For reference, the Kilo-Degree Survey achieved a precision of $\sigma_\mathrm{z}<0.02(1+z)$ per \citet{2019MNRAS.485..498M} and the Dark Energy Survey $\sigma_\mathrm{z}<0.006(1+z)$ per \citet{2020PhRvD.102b3509A} and \citet{2025PhRvD.112h3535A}. \textit{Euclid} is expected to improve upon these uncertainties \citep{Euclid,2025A&A...697A...1E}. Given the above challenges and opportunities, an unbiased cosmological inference from CC requires an accurate modelling of the cluster bias, photometric redshift uncertainties, small-scale clustering, and knowledge of the purity of the sample.

The paper is structured as follows: In Sect. \ref{sec:method}, we detail the methodology we apply for the measurement and modelling of the redshift-space 2PCF in the \code{PINOCCHIO} mock catalogues. In Sect. \ref{sec:results} we present our comparison of the model predictions to the measured 2PCF. In Sect. \ref{sec:conclusions}, we highlight our main findings and provide an outlook for future studies. Throughout this paper, we assume the best-fit {\it Planck} 2013 lambda cold dark matter ($\Lambda$CDM) cosmological parameters \citep[][Table 2]{2014A&A...571A..16P}, which were used in the construction of the \code{PINOCCHIO} mock catalogues.

\section{Method} \label{sec:method}

\subsection{Simulated cluster catalogues}

To quantify the accuracy of the \citet{1994MNRAS.267.1020P}, \citet{2004PhRvD..70h3007S}, and \citet{2010PhRvD..82f3522T} models, we tested their predictions with simulated cluster catalogues. We used 1000 past light cones simulated with the semi-analytic \code{PINOCCHIO} algorithm \citep{pinocchio,2017MNRAS.465.4658M}. A past light cone is a mock catalogue constructed as an observer would see it: objects are selected along the observer’s past line of sight, with their positions and redshifts corresponding to different cosmic times rather than to a single simulation snapshot. Thus, each light cone represents one mock realisation of the survey volume, including the redshift evolution of the halo population across the analysed range. The 1000 light cones are generated with independent initial conditions, rather than from repeated measurements of the same underlying light cone. Each realisation therefore provides an independent draw of the survey geometry and selection applied to a different mock halo distribution. The code adopts third-order LPT to obtain the solution to ellipsoidal collapse. The boxes have a side length of $3870\,h^{-1}\,\mathrm{Mpc}$, contain 2160$^3$ particles and are in redshift space. A more extended summary of the technical specifications is given in \citet{2025A&A...697A.140F}. We highlight that \code{PINOCCHIO}, as an LPT-based simulation: (i) provides an approximate description of structure formation on small scales; (ii) predicts halo bias with significant systematic uncertainty; and (iii) can exhibit scale-dependent deviations from $N$-body simulations. 

\begin{figure}
\centering
\includegraphics[width=0.48\textwidth]{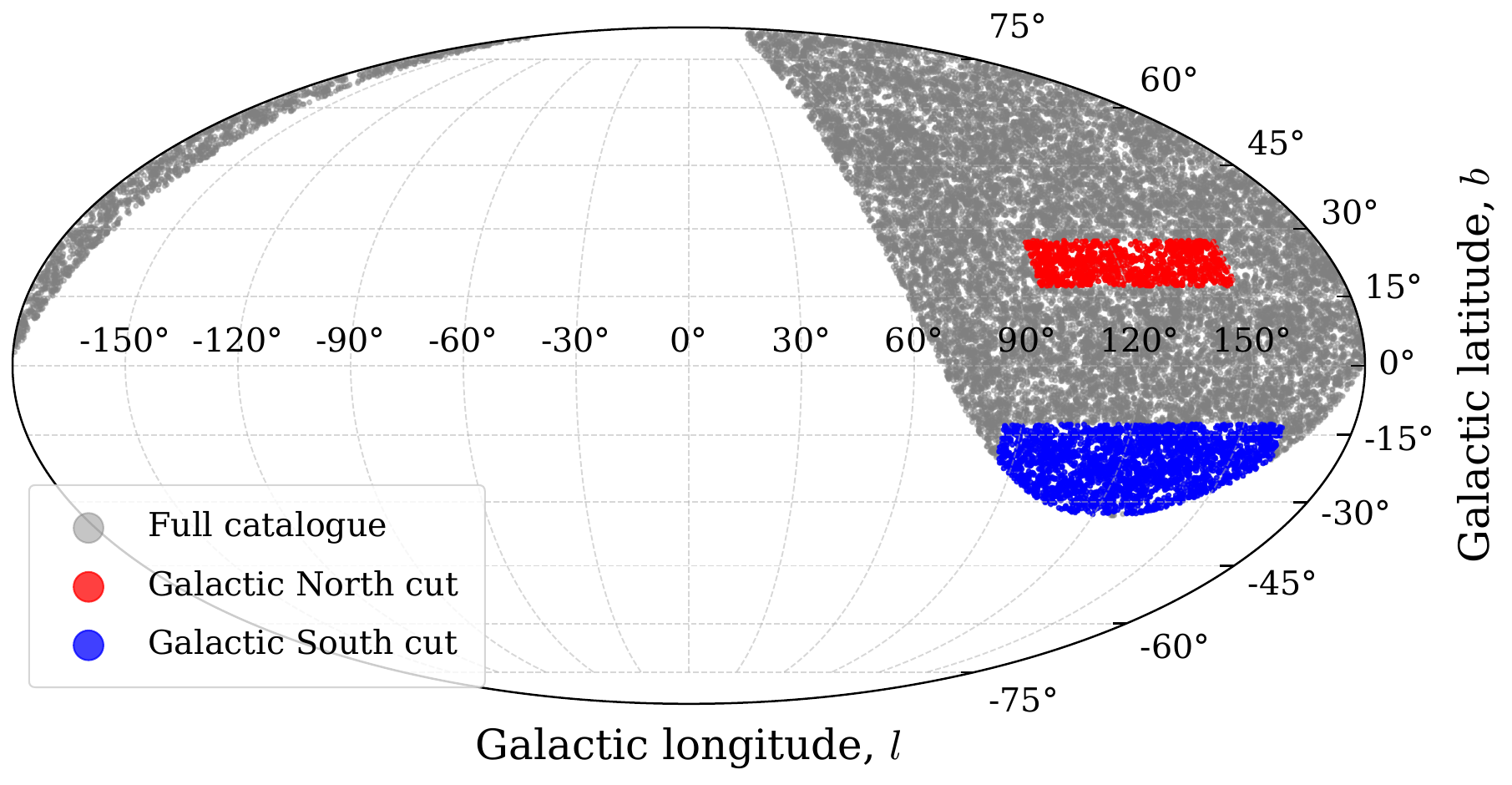}
\caption{Footprint of the \code{PINOCCHIO} catalogue used in the present analysis before (grey) and after DR1-like cuts. The red and blue patches represent DR1-like footprints in the northern Galactic hemisphere ($500$ deg$^2$) and southern Galactic hemisphere ($1400$ deg$^2$), respectively.}
\label{fig_1}
\end{figure}

The above effects may contribute to the residual discrepancies between the models and mocks and therefore to the empirical cut-off scales inferred below. The primary goal of our analysis is to define a reproducible mock-based procedure for determining such cut-off scales. The additional diagnostics presented in Appendix \ref{app:nl_clustering} are intended to clarify the likely modelling ingredients contributing to the residuals, but they do not constitute a complete physical calibration of the breakdown scale, which will be explored in future works. Furthermore, we chose \code{PINOCCHIO} to monitor the average behaviour of our inferences across multiple realisations and because \code{PINOCCHIO} provides more control over the halo mass function calibration. The original catalogues cover a sky area of $\Omega_\mathrm{sky}=10\,313\deg^2$. For the present study, we measure CC in the range $0<z<2$, impose a lower virial mass cut of $10^{14}$ $h^{-1}M_\odot$, which roughly mimics the \textit{Euclid} selection \citep{2016MNRAS.459.1764S, adam19} and isolate a DR1-like footprint as shown in Fig. \ref{fig_1}. Specifically, we assume $500$ deg$^2$ in the northern Galactic hemisphere and $1400$ deg$^2$ in the southern Galactic hemisphere. The resulting halo number densities per redshift bin are shown in Table \ref{tab:num_dens}. In the real catalogues, the southern and northern tiles will also differ in terms of sky coverage, but also in their photometric redshift uncertainties and in the selection function of the sources therein. 

In what follows we define the halo mass as a virial mass, $M_{\rm vir}$, which is the mass enclosed within a sphere of a radius, $r_{\rm vir}$, whose mean density is $\Delta_{\rm vir}(z)$ times the critical density at redshift, $z$, such that $ M_{\rm vir} = \frac{4\pi}{3}\,\Delta_{\rm vir}(z)\,\rho_{\rm c}(z)\,r_{\rm vir}^3$. When expressed in terms of $M_{\rm vir}$, the halo mass function is known to be approximately universal in redshift for standard $\Lambda$CDM cosmologies \citep{2016MNRAS.456.2486D}, which motivates our choice of mass definition throughout this work. In our catalogues, the halo masses are rescaled to follow the \citetalias{castro23} halo mass function. Details on the rescaling process can be found in \citet{2021A&A...652A..21F}. Further, we assume completeness and purity of $100\%$ across our mass ($M_\mathrm{vir}>10^{14} h^{-1}M_\odot$) and redshift ($0<z<2$) ranges. Completeness is the fraction of all existing sources within the specified mass and redshift ranges that are successfully detected by the survey. Purity is the fraction of detected sources that are clusters, rather than false detections. This is a somewhat optimistic estimate compared to the results of forecasts across $30$ deg$^2$ on the Flagship simulation, which predict a purity of 80--90\% and a completeness of more than $80\%$, with the Adaptive Matched Identifier of Clustered Objects (AMICO) described in \citet{2018MNRAS.473.5221B} and the wavelet-based algorithm PZWav described in \citet{Gonzalez2014}, which are the official methods used in \textit{Euclid}. In our analysis, the assumption of perfect completeness reduces the shot noise per bin and thus slightly increases the $f\sigma_8$ precision. The impact of assuming perfect purity depends on the nature of the contaminants. Both the assignment of clusters to the wrong redshift or mass bin and mis-identification of other sources as clusters would result in a modification of the sample's effective bias. However, the latter could also impact the sample's clustering properties more significantly. The magnitude of these effects will need to be assessed once realistic estimates of the cluster purity from the Euclid detection pipelines become available. 

We allow this mildly optimistic choice because it reduces the statistical error, thus improving the robustness of the determination of the cut-off scales, which would otherwise be completely dominated by noise. This is not critical here, as our goal is to assess systematic effects in the modelling, rather than forecast the exact precision of cosmological constraints. We also note that any incompleteness does not affect the shape of the 2PCF. Finally, we assume that projection effects (such as the contamination of optically selected cluster catalogues by structures aligned along the line of sight, which can affect the inferred cluster redshift, mass proxy, and effective bias. Therefore, we chose to modify the clustering amplitude) would at least be partially mitigated by the cluster detection and analysis pipelines.

We generated halo catalogues \citep{2025arXiv251013509F} with photometric cluster redshift uncertainties that follow a Gaussian distribution $\mathcal{N}(z^\mathrm{ob}|z, \sigma_\mathrm{z})$ with mean $z$ and standard deviation,
\begin{equation}
    \sigma_\mathrm{z} = \sigma_0\;(1+z)\,,
\end{equation}
where $\sigma_0=0.005$ for optimistic and $\sigma_0=0.01$ for realistic redshift uncertainties, as also assumed in the study of systematic effects for the full-shape analysis \citep{2025arXiv251013509F}. We also consider $\sigma_0=0$ as a reference.

\subsection{Two-point correlation function estimation}

We measured the 2PCF with its associated covariance matrices per comoving distance bin across $1000$ mock catalogues, where the random-to-data ratio is $50$. We used $15$ logarithmic radial separation bins from $r_\mathrm{min}=10$ \Mpch to $r_\mathrm{max}=150$ \Mpch in five redshift bins of width $\Delta z=0.4$ at $0<z<2$. We chose this smaller number compared to \citet{2025arXiv251013509F} to increase the S/N per radial separation bin. Throughout this study, for the measurement of the 2PCF and the clustering model prediction, we used the CosmoBolognaLib\footnote{\href{https://gitlab.com/federicomarulli/CosmoBolognaLib}{gitlab.com/federicomarulli/CosmoBolognaLib}}, a large set of free software C++/Python libraries \citep{2016A&C....14...35M}. We used the direct \citet{1993ApJ...412...64L} estimator, expressed as
\begin{eqnarray}
    \hat{\xi}_\ell(r)&=&\int_{-1}^{1}\left[\mathrm{RR}(r)+N_\mathrm{R}\mathrm{DD}(\mu,r)-2\frac{N_\mathrm{RR}}{N_\mathrm{DR}}\mathrm{DR}(\mu,r)\right]\mathcal{P}_\ell(\mu){\rm d}\mu \nonumber \\
    &\times&\frac{2\ell+1}{2\mathrm{RR}(r)} \nonumber\\
    &=& \frac{2\ell+1}{2\mathrm{RR}(r)}\left[\mathrm{RR}_\ell(r)+N_\mathrm{R}\mathrm{DD}_\ell(r)-2\frac{N_\mathrm{RR}}{N_\mathrm{DR}}\mathrm{DR}_\ell(r)\right]\,,
\end{eqnarray}
where $\ell$ is the order of the multipole ($\ell=0,2,4$ here), $\mathrm{DD}_\ell(r)$, $\mathrm{RR}_\ell(r)$, and $\mathrm{DR}_\ell(r)$ denote the Legendre-weighted data-data, random-random, and data-random pair counts at a comoving separation $r\pm \Delta r/2$, normalised by the total number of respective pairs. The number of
pairs in the random catalogue is given by $\mathrm{RR}(r)$, while $\mathcal{P}_\ell$ denotes the Legendre polynomials and $\mu$ is the cosine of the angle between the pair-separation vector and the line of sight. The number of objects in the random catalogue is given by $N_\mathrm{RR}$ and the ratio of $N_\mathrm{RR}$ over the number of objects in the data catalogue is given by $N_\mathrm{R}$. The normalisation for data–random cross-pairs is given by $N_\mathrm{DR}$. This estimator is therefore equivalent to the usual $\mu$-integrated multipole estimator in the limit of infinitesimally fine $\mu$ bins, but avoids explicitly constructing the intermediate $\xi(r,\mu)$ by computing the multipoles directly from the pair counts with no $\mu$-binning introduced. The cosine of the angle between the separation vector and the line of sight is denoted by $\mu$ such that $\mu=\cos{\theta}=r_{||}/r$, where $r$ is the absolute separation. 

In the next step, we constructed random catalogues, which contain randomly distributed clusters that serve as a reference against which clustering in the mock data is measured. To construct the random catalogues used in our clustering measurements, we subsampled a fixed fraction of clusters from each of $1000$ independent \code{PINOCCHIO} mock cluster catalogues. Specifically, from each mock we randomly selected 1 in 50 objects (i.e., a 2\% sampling rate), yielding a total of 50 times more objects in the random than in the data catalogues. This approach yields data and random catalogues that reproduce the DR1 selection sky area, without modelling the detailed angular selection or any position-dependent effect. 

\subsection{Redshift-space clustering models}

In this section, we describe the different models we test for the redshift-space 2PCF. Throughout our analysis, we restrict to models that are linear in the bias, so that, besides the cosmological parameters, they depend on a single additional parameter. We compute the theoretical 2PCF multipoles as
\begin{equation}
\xi_\ell(r)
= \frac{2\ell+1}{2}
\int_{-1}^{1} {\rm d}\mu\;\xi(r,\mu)\;\mathcal{P}_\ell(\mu) 
\quad ,\,\ell = 0,2,4\,.
\label{eq:xi_ell}
\end{equation}
We denote the linear matter auto power spectrum by $P_{\mathrm{\delta}\mathrm{\delta}}$, the velocity divergence auto power spectrum by $P_{\theta\theta}$ and their cross-power spectrum by $P_{\delta\theta}$. In all variants described below, we compute the linear matter power spectrum using \code{CAMB} \citep{2011ascl.soft02026L}. An extension of the Kaiser model to smaller scales is the dispersion model \citep{1994MNRAS.267.1020P} via
\begin{equation}
P^s(k, \mu) = D(k\mu\sigma_\mathrm{v})\;\left(1+\frac{f}{b}\mu^2\right)^2\;b^2\;P_{\mathrm{\delta}\mathrm{\delta}}(k)\,,
\end{equation}
where $s$ denotes the redshift space, $D$ an empirical damping factor, $k$ the wave number, and $b$ the linear bias. The last two terms in the above equation are the Kaiser model \citep{1987MNRAS.227....1K}. Beyond the {\it{Planck}} 2013 cosmological parameters at which we evaluate $b\sigma_8$, we assume that the velocity dispersion scale,  $\sigma_\mathrm{v}$, contains only the photometric damping contribution, expressed as
\begin{equation}
\sigma_\mathrm{v} = \frac{c\sigma_\mathrm{z}}{H(z)}\,,
\label{eq:sigmav}
\end{equation}
where $H(z)$ is the Hubble parameter because the \code{PINOCCHIO} mock catalogues do not model the physical velocity dispersion, which is, at the very least, comparable to the redshift uncertainties considered here. The models are generally insensitive to small changes to $\sigma_\mathrm{v}$. We note that the damping scale effectively captures the redshift-uncertainty contribution, while the physical velocity dispersion is expected to be subdominant.

The damping factor can be either Gaussian or Lorentzian and mimics the damping due to velocity dispersion \citep{2012MNRAS.427..327D}. Here, we use the Gaussian variant in line with the specifications of the \code{PINOCCHIO} mock catalogues via
\begin{equation}
D(k\mu\sigma_\mathrm{v}) =
\exp{\left[-\left(k\mu\sigma_\mathrm{v}\right)^2\right]}\,.
\end{equation}
The next clustering model we consider is the \citet{2004PhRvD..70h3007S} model, which accounts for the coupling between the velocity divergence and the matter density at small scales
\begin{equation}
P^s(k,\mu)=D(k\mu\sigma_\mathrm{v})\;\!\bigl[
b^2 P_{\delta\delta}(k)+2bf\,\mu^2 P_{\delta\theta}(k)+f^2\mu^4 P_{\theta\theta}(k)
\bigr]\,.
\end{equation}
Although more accurate than the dispersion model, the Scoccimarro model is still known to become inaccurate on small scales \citep{2019A&A...622A.109B}. \citet{2010PhRvD..82f3522T} proposed a more accurate perturbative model, which we refer to as TNS via
\begin{eqnarray}
P^s(k, \mu) &=& D(k\mu\sigma_\mathrm{v})\, \left[b^2 P_{\mathrm{\delta}\mathrm{\delta}}(k)+2fb\mu^2P_{\delta\theta}(k)+f^2\mu^4P_{\theta\theta}(k)\right]\nonumber \\
&+&D(k\mu\sigma_\mathrm{v})\left[C_A(k, \mu, f, b)+C_B(k, \mu, f, b)\right]\,,
\end{eqnarray}
where
\begin{eqnarray}
C_A(k, \mu, f, b) &=& b^3\;(A_2+A_4+A_6)\,,\\
C_B(k, \mu, f, b) &=& b^4\;(B_2+ B_4+ B_6+ B_8)\,,
\end{eqnarray}
and
\begin{equation}
A_2 = \mu^2\;\left[\frac{f}{b}P_{A,11}(k, \mu)+\left(\frac{f}{b}\right)^2P_{A,12}(k)\right]\,,
\end{equation}
\begin{equation}
A_4 = \mu^4\;\left(\frac{f}{b}\right)^2 \left[P_{A,22}(k)+\frac{f}{b}P_{A,23}(k)\right]\,,
\end{equation}
\begin{equation}
A_6 = \mu^6\;\left(\frac{f}{b}\right)^3 P_{A,33}(k)\,,
\end{equation}
\begin{equation}
B_2 = \mu^2\;\left[\left(\frac{f}{b}\right)^2 P_{B,12}(k)+\left(\frac{f}{b}\right)^3P_{B,13}(k)+\left(\frac{f}{b}\right)^4P_{B,14}(k)\right]\,,
\end{equation}
\begin{equation}
B_4 = \mu^4\;\left[\left(\frac{f}{b}\right)^2 P_{B,22}(k)+\left(\frac{f}{b}\right)^3P_{B,23}(k)+\left(\frac{f}{b}\right)^4P_{B,24}(k)\right]\,,
\end{equation}
\begin{equation}
B_6 = \mu^6\;\left[\left(\frac{f}{b}\right)^3 P_{B,33}(k)+\left(\frac{f}{b}\right)^4P_{B,34}(k)\right]\,,
\end{equation}
\begin{equation}
B_8 = \mu^8\;\left(\frac{f}{b}\right)^4P_{B,33}(k)\,,
\end{equation}
where we have omitted explicitly the dependence on $\mu$ for brevity and $P_{xx}$ represent the 1-loop power spectra computed using standard perturbation theory in the \code{CPT} library \citep{2008ApJ...674..617T}. The dispersion, Scoccimarro, and TNS models produce very similar predictions in the presence of photometric redshift uncertainties. In Fig. \ref{fig_3}, we present a comparison of all three models for fixed parameters. As the redshift uncertainty increases, the photometric redshift damping suppresses small-scale features, reducing the differences between model predictions. In all cases, the three models do not produce substantially different goodness-of-fit metrics based on our mock tests for a cut-off scale of $10\,h^{-1}\,\mathrm{Mpc}$ when photometric redshift uncertainties are present. This result is in accordance with the behaviour shown in Fig. \ref{fig_3}. We therefore proceeded with our analysis using the dispersion model, which is the computationally least expensive one. To compare the redshift-space power-spectrum models to the measured configuration-space multipoles, we transformed the model $P^s(k,\mu)$ into $\xi_\ell(r)$. We first computed the power-spectrum multipoles via
\begin{equation}
P_\ell(k)=\frac{2\ell+1}{2}\int_{-1}^{1}\mathrm{d}\mu\;P^s(k,\mu)\,\mathcal{P}_\ell(\mu)\,.
\end{equation}
We then obtained the configuration-space multipoles, $\xi_\ell(r)$ from $P_\ell(k)$, using an FFTLog-based Hankel transform (spherical-Bessel transform). We assumed a linear bias, $b$, given by the prescription,
\begin{equation}
b = b_\mathrm{eff}(z) = \frac{\int_{M_\mathrm{min}}^{M_\mathrm{max}} {\rm d}M\;b(M, z)\;\Phi(M, z)}{\int_{M_\mathrm{min}}^{M_\mathrm{max}} {\rm d}M\;\Phi(M, z)}\,,
\label{eq:bias}
\end{equation}
where $\Phi(M, z)$ is the halo mass function and $b(M, z)$ is the halo bias. In this work, we considered the prescriptions of \citetalias{tinker} (halo mass function), \citetalias{castro23} (halo mass function), and \citetalias{2024AA...691A..62E} (halo bias). We adopted the virial mass integration limits $M_\mathrm{min} = 10^{14}h^{-1}M_\odot$ and $M_\mathrm{max} = 10^{17}h^{-1}M_\odot$, which contain all the halos in the mock catalogues above the mass threshold. The effective bias can be integrated over redshift to account for the non-negligible bin width, yet the result is effectively the same as computing the bias at the centre of the redshift bin. Both photometric redshift uncertainty scenarios impacting the halo mass function yield almost the same effective bias.

\subsection{Likelihood}

We assumed a Gaussian likelihood centred around the measurement of the 2PCF for each catalogue. We used $1000$ \code{PINOCCHIO} light cones to estimate the 2PCF covariance. Performing a Kolmogorov-Smirnov test, we found that the Gaussian likelihood is a sufficient choice for all multipoles. We set $\theta=\{f\sigma_8\}$ as the parameter of interest, $\xi^\mathrm{m}_\ell(r)$ as the clustering model, and $\xi^\mathrm{d}_\ell(r)$ is the measured 2PCF. The likelihood is then expressed as
\begin{equation}
 \mathcal{L}(\theta) =
 \frac{1}{\sqrt{|2\pi \tens{C}|}}
 \exp\!\left[
 -\frac{1}{2}\,
 \Delta(\theta)^\top \tens{C}^{-1}\Delta(\theta)
 \right]\,,
\end{equation}
where $\Delta(\theta)$ is the residual data vector entering the fit, constructed by concatenating the residuals of all retained radial-separation bins and multipoles. Its components are
\begin{equation}
\Delta_{(\ell,i)}(\theta)
=
\xi^\mathrm{d}_\ell(r_i)
-
\xi^\mathrm{m}_\ell(r_i;\theta)\,,
\end{equation}
for $\ell=0,2$ and $r_i>r_{\rm min,\ell}$, where $i$ labels the radial-separation bins and $r_{\rm min,\ell}$ denotes the minimum scale retained for the multipole of an order of $\ell$. 

While we find that the 2PCF distribution per radial separation bin is well approximated by a Gaussian distribution, we propose the use of a Student's-t distribution since the covariance matrix is estimated numerically from a finite set of mocks, in the regime where the Gaussian approximation may fail in future analyses \citep{2022MNRAS.510.3207P}. The Hartlap correction \citep{2007A&A...464..399H} is around $1$ because we used 1000 light cones to estimate the covariance, but $45$ 2PCF bins and one inferred parameter. 

We find that the cross-correlations between redshift bins are negligible for all multipoles, regardless of the presence of redshift-space distortions or redshift uncertainties. This is in line with the findings by \citet{2025arXiv251013509F} and \citet{2024A&A...683A.253E} for the monopole. Nevertheless, in all subsequent analyses, we used the full covariance matrix. In the present study, we find that the assumption of independence between redshift bins holds also for the quadrupole and hexadecapole, although we excluded the latter from our analysis. This is likely because the redshift bin width we consider is greater than $1\,h^{-1}$Gpc, whereas the redshift uncertainties at worst correspond to 80 $h^{-1}$Mpc. It is therefore safe to combine the bins. However, since our goal here has been to assess the clustering model accuracy per bin rather than perform a forecast, we carried out our inferences in individual bins.

We fixed the BAO parameters to $\alpha_\parallel = 1$ and $\alpha_\perp = 1$ \citep[][equations 4--5]{2014MNRAS.439...83A}, and $\sigma_v$ to the value derived from Eq. \eqref{eq:sigmav}. These modelling choices help ensure that the derived cut-off scales are driven by $f\sigma_8$. The former choice is driven by the weak constraints on $\alpha_\parallel, \alpha_\perp$ when assuming uniform priors across all model parameters. We assess the convergence of our $f\sigma_8$ Markov chain Monte Carlo chains by computing the effective sample size (ESS) for $f\sigma_8$. We find an ESS of 1300--1800, consistent with negligible autocorrelation and near-independent samples. Performing a Gelman--Rubin test \citep{1992StaSc...7..457G}, we find $R<1.11$ at most. We conclude that all chains are fully converged. 

\subsection{Cut-off scale selection criteria}\label{sec:criteria}

Given DR1's large uncertainties, our plan is to first assess model adequacy through goodness-of-fit of the model considering a common cut-off scale across multipoles, which is allowed to vary with redshift. In this case, we require that $|\chi_{\rm red}^2-1|\leq \sigma(\chi_{\rm red}^2)$. For $N_\mathrm{dof}$ degrees of freedom, where $N_\mathrm{dof} = N_{\rm bins}^{(0)} + N_{\rm bins}^{(2)} - 1$ (the number of monopole and quadrupole bins above the scale cuts, respectively, minus the single fitted parameter $f\sigma_8$), the expected fluctuation of the $\chi^2_\mathrm{red}$ is
\begin{equation}
        \sigma\left(\chi^2_{\rm red}\right) = \sqrt{\frac{2}{N_\mathrm{dof}}}\,.
    \end{equation}
This is the primary criterion for determining whether the data warrant increased model complexity. We also want to check whether the fitted models yield unbiased posteriors. However, given the large uncertainties in a DR1-like setting, where the $f\sigma_8$ posterior is largely prior-dominated, any point estimator will not yield a reliable test of bias, as the posteriors end up being left-truncated and heavily skewed. Consequently, restrictive requirements on bias will not apply.

Below, we further introduce a combination of more involved statistical tests to be considered in future analyses. We refer to the former and the latter as permissive and conservative tests, respectively. Given the low information content of the $f\sigma_8$ posteriors at high redshifts, we only plan to apply bias-related metrics to posteriors that are not close to uniform for the purposes of demonstration. Accordingly, we will keep the requirements for this additional, more conservative combination of tests relatively loose. Cut-off scales are generally expected to vary with redshift, redshift uncertainty, and multipole order, leading to a very large number of plausible combinations of cut-off scales. In a DR1-like setting, we will exclude the hexadecapole from our search since we have found that the information content of the $f\sigma_8$ posteriors does not change upon its inclusion, as we discuss in Appendix \ref{app:hexadecapole}. We would then perform a 2D grid search over $(r_{\min,0}, r_{\min,2})$, scanning all admissible cut-off scale pairs and selecting only the ones that satisfy our chosen statistical consistency tests. To limit the size of the parameter space, we restricted the cut-off scales to $r_{\min,0}, r_{\min,2} \in {10, 20, 30, 40, 45, 55}\,h^{-1}\mathrm{Mpc}$. We then inferred $f\sigma_8$ from the 2PCF multipoles truncated at each pair of minimum scales in this grid. 

Subsequently, we applied the following statistical requirements and conditions to each
$(r_{\min,0},r_{\min,2})$ pair, so that every multipole separation $r_\ell>r_\mathrm{min,\ell}$ will be admissible:
\begin{itemize}
    \item For the standard scores, for each catalogue, we define the standard score of the posterior median as
    \begin{equation}
        s = \frac{q_\mathrm{50} - (f\sigma_8)_{\rm true}}
                 {\sigma_{\rm post}}\,,
    \qquad
    \sigma_{\rm post} = \frac{q_\mathrm{84} - q_\mathrm{16}}{2},
    \end{equation}
    where $q_\mathrm{50}$ is the posterior median, while $q_\mathrm{84}$ and $q_\mathrm{16}$ are the 84th and 16th percentiles, respectively. We then define the distance in the bias-variance plane from a perfectly statistically calibrated model \citep{David1948PIT}, as
    \begin{equation}
        D_\mathrm{bv}^2 = \langle s \rangle^2 + \left[\sigma(s)-1\right]^2\,,
    \end{equation}
    where each catalogue yields one value of $s$, $\langle s \rangle$ is the figure of bias, and $\sigma(s)$ the spread of the standard score distribution across catalogues. For a model perfectly calibrated across $N_{\mathrm{cat}}$ catalogues such that $s\sim \mathcal{N}(0,1)$, finite-sample fluctuations give $\sigma(\langle s \rangle) \approx 1/\sqrt{N_{\mathrm{cat}}}$ and
    $\sigma\left[\sigma(s)-1\right] \approx 1/\sqrt{2N_{\mathrm{cat}}}$. This implies a characteristic total noise level of $\sigma(D_\mathrm{bv}) = \sqrt{3/(2N_{\mathrm{cat}})}$. This condition ensures that the resulting posteriors are statistically consistent with their stated uncertainties. Following the above, a $3\sigma$ envelope corresponds to $D_\mathrm{bv} \approx 0.4$ for $N_{\mathrm{cat}}=100$. Therefore, we adopt the following requirement,
    \begin{equation}
    D_\mathrm{bv} \lesssim 0.4\,.
    \label{eq:D_bv}
    \end{equation}
    Here, we find that $N_\mathrm{cat}=100$ mock catalogues are sufficient to ensure statistical stability across all statistical tests. Increasing the number of catalogues beyond this value does not change our conclusions for any of the tests presented in this section.
    \item For the highest posterior-density coverage test, for each nominal credibility level, $\alpha$, among $\{0.10,0.30,0.50,0.68,0.90,0.95\}$, we compute the empirical highest-posterior-density (HPD) coverage $\tens{C}_\mathrm{HPD}(\alpha)$ across the catalogues. Under perfect calibration, $\tens{C}_\mathrm{HPD}(\alpha)$ fluctuates as a binomial estimator with sampling uncertainty of $\sigma_\alpha = \sqrt{\alpha(1-\alpha)/N_{\mathrm{cat}}}$. The worst case occurs near $\alpha\simeq0.68$, giving a $3\sigma$ envelope of $\sim 0.14$ for $N_{\mathrm{cat}}=100$. Therefore, we require
    \begin{equation}
        \max_\alpha |\tens{C}_\mathrm{HPD}(\alpha)-\alpha| \lesssim 0.14\,,
        \label{eq:hpd}
    \end{equation}
    ensuring that no credibility level is severely miscalibrated beyond finite-sample fluctuations. This requirement is analogous to demanding that the $68\%$ credible interval contain the truth in $68\%$ of realisations, but it is stricter because it tests calibration simultaneously at multiple $\alpha$ values and uses HPD intervals, which remain robust for highly skewed or truncated posteriors.
    \item The monotonicity condition states that if a cut-off scale pair is accepted, all larger scale cut-offs must also meet the above requirements.
    \item In the elimination of unstable scales, cut-off pairs that satisfy the tests, but end up yielding systematically larger $D$ values than all larger scales are discarded as a conservative safeguard, independently of whether the difference is driven by noise or by residual systematics and to what extent.
    \item To ensure catalogue-level robustness:, all criteria must be met across 100 catalogues, beyond which our results do not change substantially.
    \item To achieve the goodness-of-fit, we require that $|\chi_{\rm red}^2-1|\leq 0.5\,\sigma(\chi_{\rm red}^2)$ for the mean $\chi_{\rm red}^2$ across 100 catalogues. This test ensures that the $\chi^2_\mathrm{red}$ distribution across catalogues when fitting the monopole and quadrupole jointly will be well-centred around unity.
\end{itemize}
Taken together, the standard-score distance $D$ (including its figure-of-bias component $\langle s\rangle$), the HPD coverage criterion, and the $\chi^2_\mathrm{red}$ requirement provide a goodness-of-fit metric for the statistical calibration of the model across the ensemble of catalogues. For the inference of $f\sigma_8$ for the above tests, we split the halo catalogue into five redshift bins of width $\Delta z=0.4$ at $0<z<2$. We compute the covariance matrix, $C$, for each redshift bin across the $1000$ \code{PINOCCHIO} light cones.

\section{Results} \label{sec:results}

\begin{table}[t]
\caption{\label{tab:num_dens}Halo number densities in the mock DR1-like footprint.}
\centering
\begin{tabular}{ccc}
\hline\hline
Redshift bin & $N_{\rm halos}$ &
$\bar n~\left[h^{3}\,{\rm Mpc}^{-3}\right]$ \\
\hline
$0.0 < z < 0.4$ & 4000     & $1.47\times10^{-5}$ \\
$0.4 < z < 0.8$ & $10\,000$ & $7.72\times10^{-6}$ \\
$0.8 < z < 1.2$ & 6000     & $2.64\times10^{-6}$ \\
$1.2 < z < 1.6$ & 2000     & $6.89\times10^{-7}$ \\
$1.6 < z < 2.0$ & 500      & $1.55\times10^{-7}$ \\
\hline
\end{tabular}
\tablefoot{The number densities are calculated per redshift shell assuming a uniform radial selection.}
\end{table}

\begin{table}[!ht]
\caption{\label{tab:results}Reduced $\chi^2$ values for the dispersion-model fits.}
\centering
\renewcommand{\arraystretch}{1.2}
\begin{tabular}{ccc}
\hline\hline
Redshift bin & $\sigma_0 = 0.005$ & $\sigma_0 = 0.01$ \\
\hline
$0.0 < z < 0.4$ & $1.07 \pm 0.35$ & $1.09 \pm 0.38$ \\
$0.4 < z < 0.8$ & $1.22 \pm 0.37$ & $1.12 \pm 0.39$ \\
$0.8 < z < 1.2$ & $1.78 \pm 0.51$ & $1.46 \pm 0.42$ \\
$1.2 < z < 1.6$ & $1.65 \pm 0.54$ & $1.45 \pm 0.40$ \\
$1.6 < z < 2.0$ & $1.20 \pm 0.44$ & $1.14 \pm 0.37$ \\
\hline
\end{tabular}
\tablefoot{The values are obtained by fitting the dispersion model to 100 catalogues using the cut-off scales reported in Sect.~\ref{sec:results}. For the relevant number of degrees of freedom, the expected value is $\chi_\mathrm{red}^2=1.00\pm0.26$.}
\end{table}

\begin{figure*}
\centering
\includegraphics[width=\textwidth]{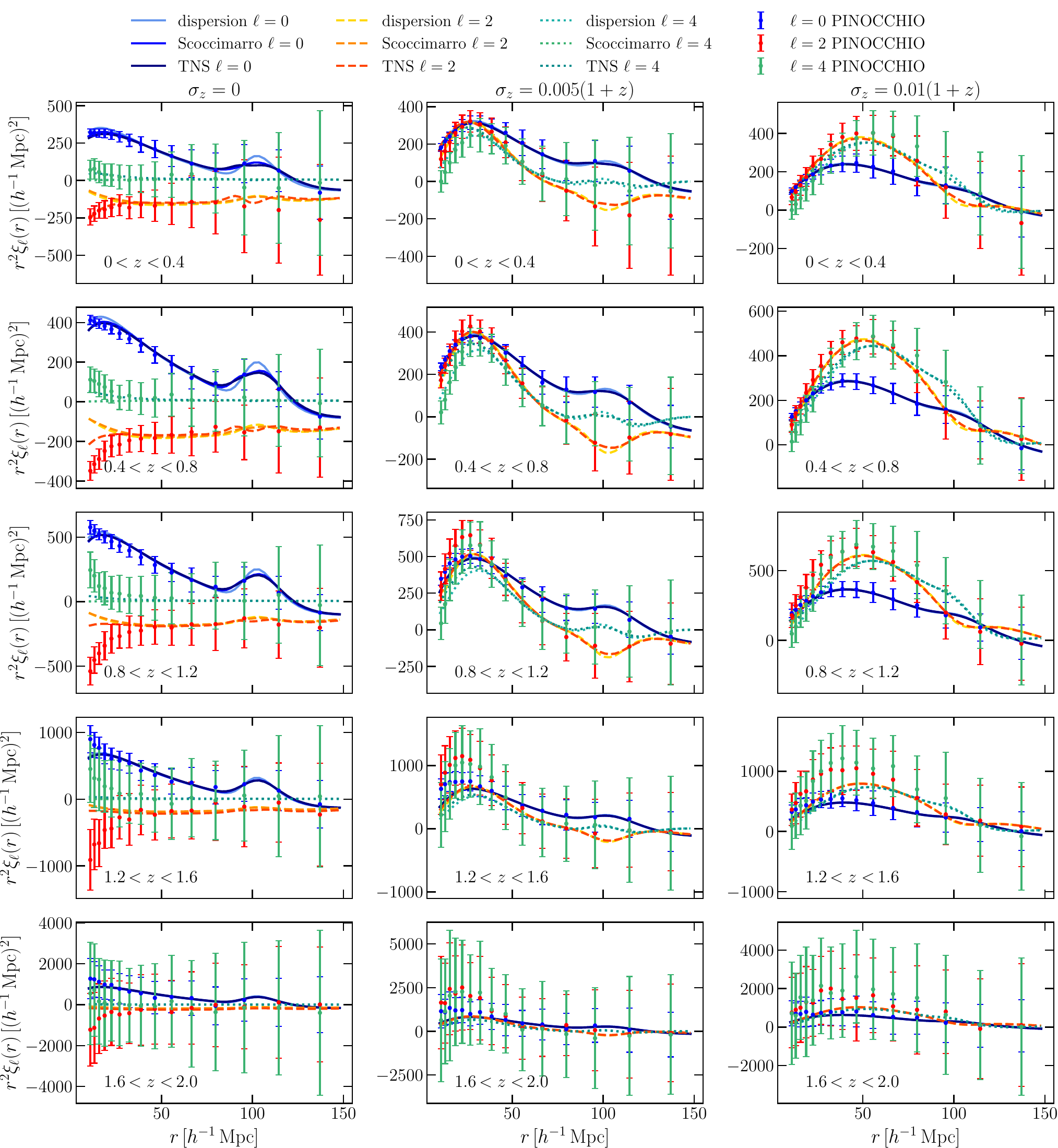}
\caption{Comparison of the dispersion, Scoccimarro, and TNS models with model parameters fixed at the ground truth of the \code{PINOCCHIO} mocks, for different redshift uncertainties (\emph{left}: none, \emph{middle}: optimistic, \emph{right}: realistic) as a function of redshift (vertically) and multipole order. The points indicate the mean 2PCF across 100 \code{PINOCCHIO} realisations and the error bars indicate the errors assumed for the DR1-like samples.}
\label{fig_3}
\end{figure*}

Using our main, permissive set of statistical tests, we find that the dispersion model constitutes a sufficient description of the data at all redshifts. Quantitatively, this translates to a $\chi^2_\mathrm{red}$ consistent with unity for $r_\mathrm{min,0}=r_\mathrm{min,2}=10$ $h^{-1}$Mpc across 100 mock catalogues using the \citetalias{tinker} bias prescription. Therefore, all cut-off scales greater than $10$ $h^{-1}$Mpc can be considered admissible. Given the large DR1 uncertainties, the inferred $f\sigma_8$ posteriors are prior-dominated. As a result, they can be left-truncated at zero and highly skewed. For such posteriors, point-estimate biases are unreliable diagnostics, as they primarily reflect the prior truncation rather than systematic model error. The only meaningful statement for uninformative posteriors is that the posteriors are consistent with the true values: across all redshift bins and mock catalogues, the true $f\sigma_8$ values fall within the credible intervals and in high-density regions of the posterior distributions. In the lower redshift bins where posteriors are more informative, the bias is less than $0.2\,\sigma$ across redshift bins for all redshift uncertainty scenarios. In the reference case of $\sigma_\mathrm{z}=0$, the cut-off scales are generally higher, since the dispersion model is not appropriate for small redshift uncertainties \citep{karcher}. The above trends are in line with the visual trends in Fig. \ref{fig_3}. We show the resulting $\chi_\mathrm{red}^2$ in Table \ref{tab:results}.

The use of the \citetalias{2024AA...691A..62E} bias prescription leads to mildly higher cut-off scales. This does not indicate that \citetalias{tinker} is inherently more accurate, but simply reflects that the semi-analytic \code{PINOCCHIO} clustering only approximates the $N$-body clustering on which the \citetalias{2024AA...691A..62E} bias was calibrated \citep{2013MNRAS.433.2389M,2017MNRAS.465.4658M,2019MNRAS.482.1786L,2019MNRAS.485.2806B}. Accordingly, we want to adopt the \citetalias{tinker} bias for the remainder of the analysis. We can verify that this choice does not introduce model mis-specification by fixing $f\sigma_8$ to its ground truth at the selected scales and recovering $b\sigma_8$ posteriors consistent with the \citetalias{tinker} prediction. In a forthcoming work, where higher fidelity non-linear mocks will be available, we will be able to employ physical priors on $b\sigma_8$ centred on these values, with variances informed by the mass–observable uncertainties expected for \textit{Euclid} \citep[section 3 in][]{2016MNRAS.459.1764S} and by non-linear halo-bias prescriptions. The higher fidelity mocks should model structure formation using an $N$-body prescription.

\begin{figure*}
\centering
\includegraphics[width=0.95\textwidth]{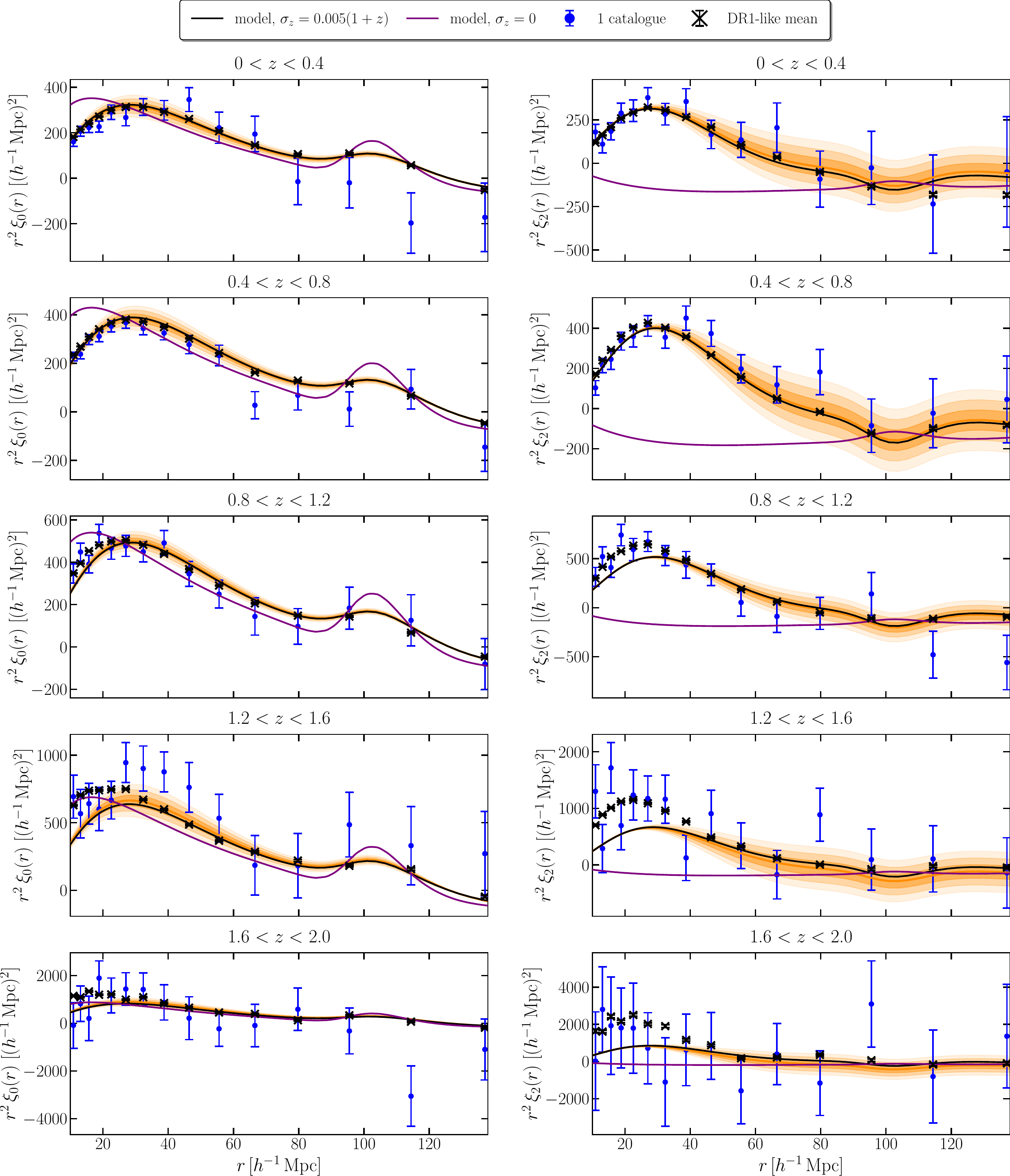}
\caption{Monopole and quadrupole of the redshift-space 2PCF for $\sigma_\mathrm{z}=0.005(1+z)$ as a function of redshift. The purple and black lines represent the redshift-space model prediction for $\sigma_\mathrm{z}=0$ and $\sigma_\mathrm{z}=0.005(1+z)$, respectively. In blue, we show a single randomly drawn catalogue, with its $1\sigma$ uncertainty, as a reference for the size of the error bars. The black crosses represent the mean of the 2PCF (DR1-like mean) across 100 catalogues, with the error on the mean. We show the fitted model with $1-2-3\sigma$ uncertainties across realisations in orange.}
\label{fig_7}
\end{figure*}

\begin{figure*}
\centering
\includegraphics[width=0.95\textwidth]{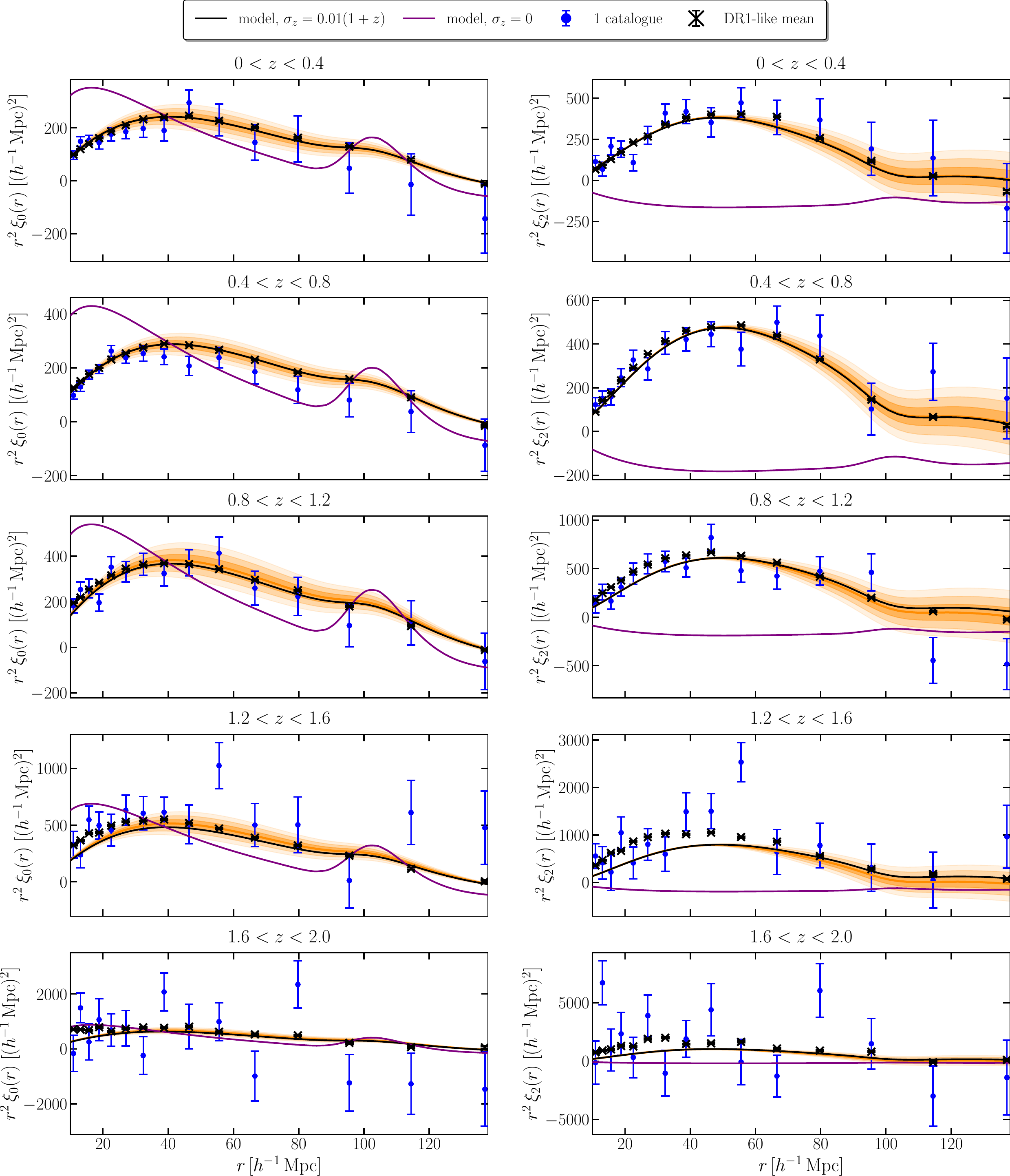}
\caption{As in Fig. \ref{fig_7}, but for $\sigma_\mathrm{z}=0.01(1+z)$.}
\label{fig_8}
\end{figure*}

\begin{figure*}
\centering
\includegraphics[width=\textwidth]{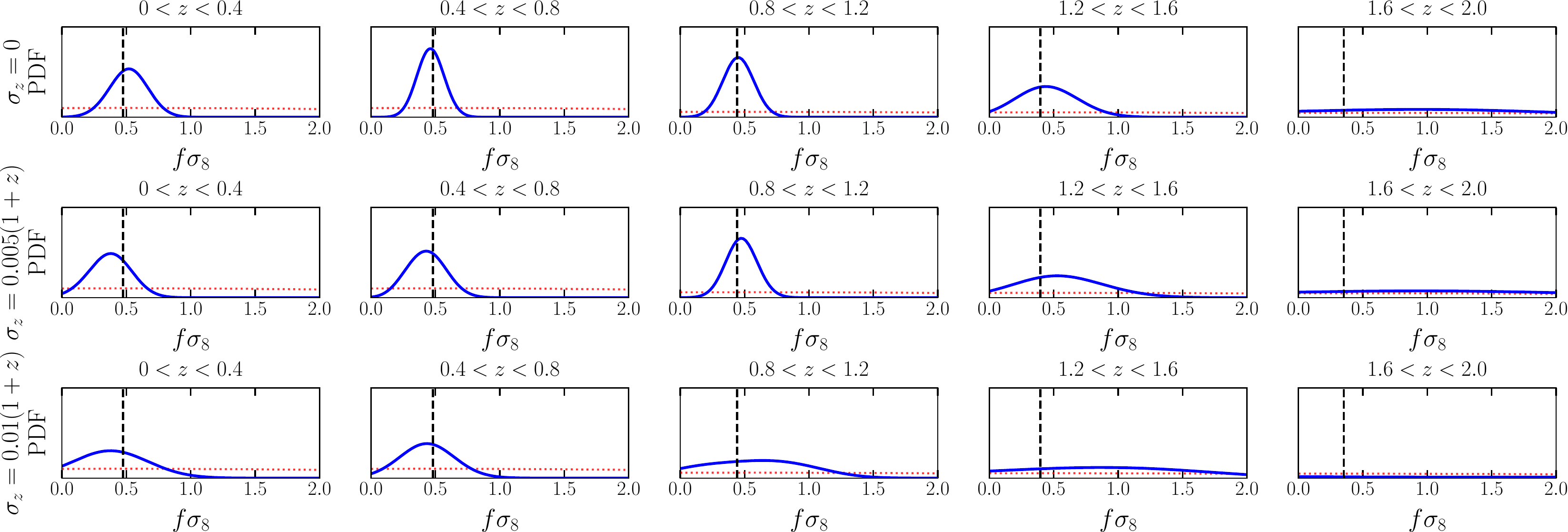}
\caption{Posterior of $f\sigma_8$ demonstrating the average behaviour of a DR1-like sample as a function of the redshift uncertainty and redshift bin. The dashed black line indicates the ground-truth $f\sigma_8$ in each bin and the red dotted line indicates the prior.}
\label{fig_5}
\end{figure*}

The corresponding fitted models for $\sigma_\mathrm{z} = 0.005(1+z)$ and $\sigma_\mathrm{z} = 0.01(1+z)$ are shown in Figs. ~\ref{fig_7} and \ref{fig_8}, respectively. Visually comparing the single-catalogue 2PCF to the model suggests that within DR1-like uncertainties, the selected cut-off scale can be treated as approximately constant with multipole order and redshift within DR1-like uncertainties. We present the typical $f\sigma_8$ posterior across 100 \code{PINOCCHIO} mocks in Fig.~\ref{fig_5}. We obtained these values by running the inference on individual mocks and then combining the posteriors resulting from the 100 realisations as independent. The priors are centred on the truth with a standard deviation of $3$, bounded below by zero and above by a bin-dependent maximum to facilitate convergence of the chains in this noisy setting. The low constraining power of DR1 on $f\sigma_8$ is unsurprising and this is very the reason we chose to use $f\sigma_8$ as a diagnostic for determining the cut-off scales: $f\sigma_8$ is highly sensitive to the small-scale quadrupole, where most of the information resides \citep[figure 2 in][]{2021JCAP...10..044F}. Therefore, the $f\sigma_8$ inference presented here will be useful also in view of future joint analyses combining cluster clustering with cluster counts, galaxy clustering, and weak lensing. Parameters whose constraining power comes from larger scales, such as $\Omega_\mathrm{m}$ and $\sigma_8$ \citep{2025arXiv251013509F}, are expected to yield tighter constraints. We emphasise that this analysis is not a forecast. Rather, it represents a best-case scenario for $f\sigma_8$ constraints under our conservative modelling of halo completeness and bias. In this analysis, we only sought to infer $f\sigma_8$, yielding 1D posteriors. In future analyses involving marginalisation over higher dimensional parameter spaces, prior volume effects can introduce biases, especially when the constraining power on cosmological parameters is low \citep[e.g.][]{2023MNRAS.525.6336H,2023JCAP...12..025M,2025JCAP...01..138M}. Potential solutions to be considered are summarised in \citet{2025arXiv250909562T} and an analysis of this effect in a \textit{Euclid}-like setting will be presented in Euclid Collaboration: Moretti et al. (in prep.).

The more conservative set of tests yields generally higher cut-off scales. For $\sigma_\mathrm{z}=0.005(1+z)$ we obtained $(r_\mathrm{min,0}, r_\mathrm{min,2})=(30,10), (20, 10),(20,30),(30,20)$ $h^{-1}$ Mpc for the first four redshift bins, respectively. For $\sigma_\mathrm{z}=0.01(1+z)$ we obtain $(r_\mathrm{min,0}, r_\mathrm{min,2})=(30,10), (10, 10),(20,30)$ $h^{-1}$ Mpc for the first three redshift bins, respectively. At higher redshifts, the $f\sigma_8$ posteriors become fairly uninformative, therefore traditional estimators like the standard score become biased. We provide further quantification of this secondary line of tests proposed to guide application to future data in Appendix \ref{app:stat_req} and Appendix \ref{app:nl_clustering}.  

Comparing the ensemble-mean 2PCF across the \code{PINOCCHIO} mocks with the corresponding true model predictions in the presence of cluster redshift uncertainties in Fig. \ref{fig_3} reveals a clear trend that will be relevant for future DRs: the quadrupole starts to deviate from the mean model at increasingly larger cut-off scales. Moreover, the scale at which this deviation becomes apparent grows with redshift and order of multipole. This behaviour might reflect that non-linearities propagate to larger scales when photometric redshifts are present due to inaccuracies of the linear halo bias, as suggested by \citet{2025arXiv251013509F} who previously reported this effect for the monopole. Here we find that it is significantly more pronounced for the quadrupole and even more so for the hexadecapole. Coherent deviations from the model across adjacent radial separation bins are expected because the measurements are correlated.

Below, we show how we can quantify the above behaviour by comparing the mean 2PCF to the dispersion model with linear halo bias. For optimistic redshift uncertainties at $0.8<z<1.2$, it can be seen that the mean monopole and quadrupole begin to deviate from the model at $r=20$ $h^{-1}$ Mpc and $r=35$ $h^{-1}$ Mpc, respectively. At $1.2<z<1.6$, these deviation scales shift to $r=35$ $h^{-1}$ Mpc for the monopole and $r=50$ $h^{-1}$ Mpc for the quadrupole. For more realistic redshift uncertainties, the deviations at $0.8<z<1.2$ occur at $r=30$ $h^{-1}$ Mpc (monopole) and $r=55$ $h^{-1}$ Mpc (quadrupole), while the cut-off scale for $1.2<z<2$ is extended to even larger scales. Overall, the behaviour of the required cut-off scales will depend jointly on multipole order, shot noise per bin (Table~\ref{tab:num_dens}), redshift, and redshift uncertainty for future, higher S/N DRs. Because these effects interact in a non-linear and redshift-dependent way, formulating a simple analytical expectation for the cut-off scales is not feasible. The aforementioned scales are derived from the mean behaviour of the sample, which can be characterised to higher precision than the behaviour of single catalogues due to the smaller 2PCF error bars. For our main analysis in Sect. \ref{sec:criteria}, we investigate the accuracy of the model against single 2PCFs per mock catalogue rather than against the mean 2PCF.

At low redshift, we find that redshift-dependent scale cuts can mitigate residual modelling inaccuracies associated with the choice of clustering model in the absence of significant photometric redshift uncertainties and with the semi-analytic nature of the PINOCCHIO mocks (see the comparison of the dispersion and TNS models for the quadrupole for $\sigma_\mathrm{z}=0$ in Fig. \ref{fig_3}). When significant photometric redshift uncertainties are included, the differences among the clustering models become largely suppressed at low redshifts. At higher redshift, however, a discrepancy between mocks and models may arise from failure of the linear halo bias model combined with photometric redshift smearing (as discussed in more detail below), rather than failure of the clustering model. In contrast, for $\sigma_\mathrm{z}=0$, the scale at which discrepancies appear remains relatively more constant with redshift. 

Specifically, photometric redshift uncertainties smooth the halo field along the line of sight. In Fourier space, for a fixed smoothing scale, this appears as a scale- and angle-dependent damping of each mode, rather than as mode coupling in the strict Fourier-space sense. In the configuration space, the same operation corresponds to a convolution along the line of sight, which mixes contributions from different radial separations. Residuals generated by non-linear effects can therefore project differently onto the observed multipoles once the field is smeared by photometric redshift uncertainties. At the same time, cluster bias increases with redshift at fixed mass threshold, so that any higher order bias contributions might become more important for the high-redshift bins. We therefore interpret the redshift dependence of the residuals as the result of a combination of non-linear tracer modelling and radial mixing induced by photometric redshift uncertainties, rather than as a consequence of non-linear clustering at high redshift alone. Ignoring such effects may introduce biases, some of which can be seen in the conservative test shown in Fig. \ref{fig_10}. Consequently, mitigating high-redshift mis-specification in future DRs requires improvements in the overall clustering model, without which the $1.2<z<2$ bins might require more conservative scale cuts with the current modelling. While the comparison of the mean 2PCF with the models highlights the scales where discrepancies begin to appear, the final cuts are determined by the statistical diagnostics applied to both the ensemble of DR1-like catalogues and the individual realisations, whose outcomes are influenced by the DR1-like noise level. Therefore, within the DR1-like statistical uncertainties considered here, these effects do not lead to a statistically significant failure of the validation criteria adopted in this work.

\section{Conclusions}\label{sec:conclusions}

We developed and validated an inference framework to constrain cosmological parameters from the multipoles of the two-point cluster correlation function. This work represents a key step toward preparing the \textit{Euclid} pipelines for cosmological analyses based on CC. Using semi-analytic mock datasets with a \textit{Euclid} DR1-like footprint, we applied the framework to infer $f\sigma_8$ and to determine the conditions under which redshift-space clustering models can yield accurate results on real data based on the requirements in Sect. \ref{sec:criteria}. Given DR1's large uncertainties, we employed a set of permissive statistical tests and a suite of conservative statistical tests to guide the determination of cut-off scales in higher S/N DRs in the future. The former is based on goodness-of-fit, while the latter additionally imposes stricter criteria for unbiasedness, statistical calibration, and coverage across all mock realisations. We based our main conclusions on the former.

We tested both realistic and optimistic redshift-uncertainty scenarios,
$\sigma_\mathrm{z} = 0.005(1+z)$ and $\sigma_\mathrm{z} = 0.01(1+z)$. We explored the dispersion, Scoccimarro, and TNS models, allowing the cut-off scales to vary with redshift and multipole. Our main results are summarised below.
\begin{itemize}
\item Under the DR1-like covariance and permissive validation criteria used in this work, the dispersion model yields unbiased $f\sigma_8$ inferences for the tested mock catalogues down to 10 \Mpch over the full redshift range. The resulting bias on $f\sigma_8$ is less than 0.2$\sigma$ across the informative redshift bins. This statement should be interpreted within the present mock-based validation setup. The exact thresholds considered admissible for the real data analysis will be reassessed using the final selection function, covariance, and systematic-error treatment.
\item The hexadecapole adds no significant constraining power in a DR1-like setting. 
\item The cut-off scale should be allowed to vary with multipole order for upcoming DRs.
\item For higher S/N DRs, the assumption of linear halo bias is likely to become insufficient for modelling the photometric cluster-clustering multipoles beyond $z\simeq1.2$, unless more advanced clustering models are considered.
\item As the multipole order increases, the above effect shifts the cut-off scale to higher values.
\end{itemize}
The latter behaviour is further associated with a systematic underprediction of the clustering amplitude that becomes more apparent with increasing redshift, redshift uncertainty, and multipole order, as shown in Fig.~\ref{fig_3}. The diagnostics in Appendix \ref{app:nl_clustering} suggest that non-linear halo bias contributes to this trend, but that the fitted quadratic term acts as an effective two-point coefficient rather than as a direct measurement of the physical local quadratic bias. The residuals are therefore likely to reflect a combination of non-linear tracer bias and the radial mixing induced by photometric redshift uncertainties. 

The broader issue is the need for more accurate theoretical predictions in order to extend photometric 2PCFs to smaller scales and higher redshifts for higher S/N catalogues. In this work, we adopted a linear halo-bias prescription feeding into models that incorporate non-linear structure formation to varying degrees. However, the halo field violates locality, Poisson noise, and linearity assumptions \citep[see][for a review]{2015MNRAS.450.1486A}. Hence, a dedicated calibration of these models against $N$-body simulations, including non-linear and non-local halo bias, as in \citet{2025A&A...697A...5E}, is required before halo clustering can be robustly pushed to smaller scales in \textit{Euclid} DR1 and to higher S/N datasets such as \textit{Euclid} DR2/3. Once such high-fidelity mocks become available, reliable forecasts for upcoming releases will be feasible. Finally, despite our optimistic assumptions on completeness, purity, and halo mass uncertainties (i.e. satisfactory conditions for the purposes of this analysis), we recovered only weakly informative posteriors on $f\sigma_8$. These results therefore represent a best-case scenario for DR1 and highlight the need for further improvements in modelling ahead of future \textit{Euclid} DRs.

Owing to the effects discussed above, we expect the minimum cut-off scales in future DRs to become more restrictive unless these modelling improvements are implemented. Assuming Poisson statistics, the 2PCF error bars are expected to decrease by roughly a factor of 2 in \textit{Euclid} DR2 and by a factor of 3 in \textit{Euclid} DR3 under the specifications considered here.  As shown in Fig.~\ref{fig_3}, the higher S/N in the 2PCF will push the empirically required cut-off scales to larger values unless a more accurate halo-bias prescription is adopted. Although the cosmological constraining power will increase accordingly \citep{2025arXiv251013509F}, exploiting the full potential of the \textit{Euclid} mission will require an improved modelling of small-scale clustering in the presence of photometric redshift uncertainties. To this end, future works should rely on state-of-the-art simulations \citep{2025A&A...697A...5E} to calibrate the underlying non-linear and non-local halo-bias behaviour and thereby maximise the scientific return of the upcoming DRs.

\begin{acknowledgements}

This work was supported by STFC through Imperial College Astrophysics Consolidated Grant ST/W000989/1. ET acknowledges support from the FoNS Researcher Mobility Grant 2025. This research utilised the HPC facility supported by the Research Computing Service at Imperial College London. \AckEC

\end{acknowledgements}

\bibliographystyle{aa}
\bibliography{main}

\begin{appendix}
\nolinenumbers
\section{Test results}\label{app:test_results}
In this section, we provide a quantification of the conservative diagnostics in Sect. \ref{sec:criteria} ($D_\mathrm{bv}$, $\max_\alpha |\tens{C}_\mathrm{HPD}(\alpha)-\alpha|$, $\chi_\mathrm{red}^2$) and present additional investigations on the constraining power of the hexadecapole and the validity of our pipeline. For brevity, we only show the cases with redshift uncertainties, as they are the focus of the present study. In the case of the $\chi_\mathrm{red}^2$ distributions, where there is one distribution for each pair of $(r_\mathrm{min,0}, r_\mathrm{min,2})$, we will present only the results for the scales chosen in Table \ref{tab:results}.

\subsection{Validation with bias and velocity dispersion inference}\label{ssec:validation}

\begin{figure}
\centering
\includegraphics[width=0.45\textwidth]{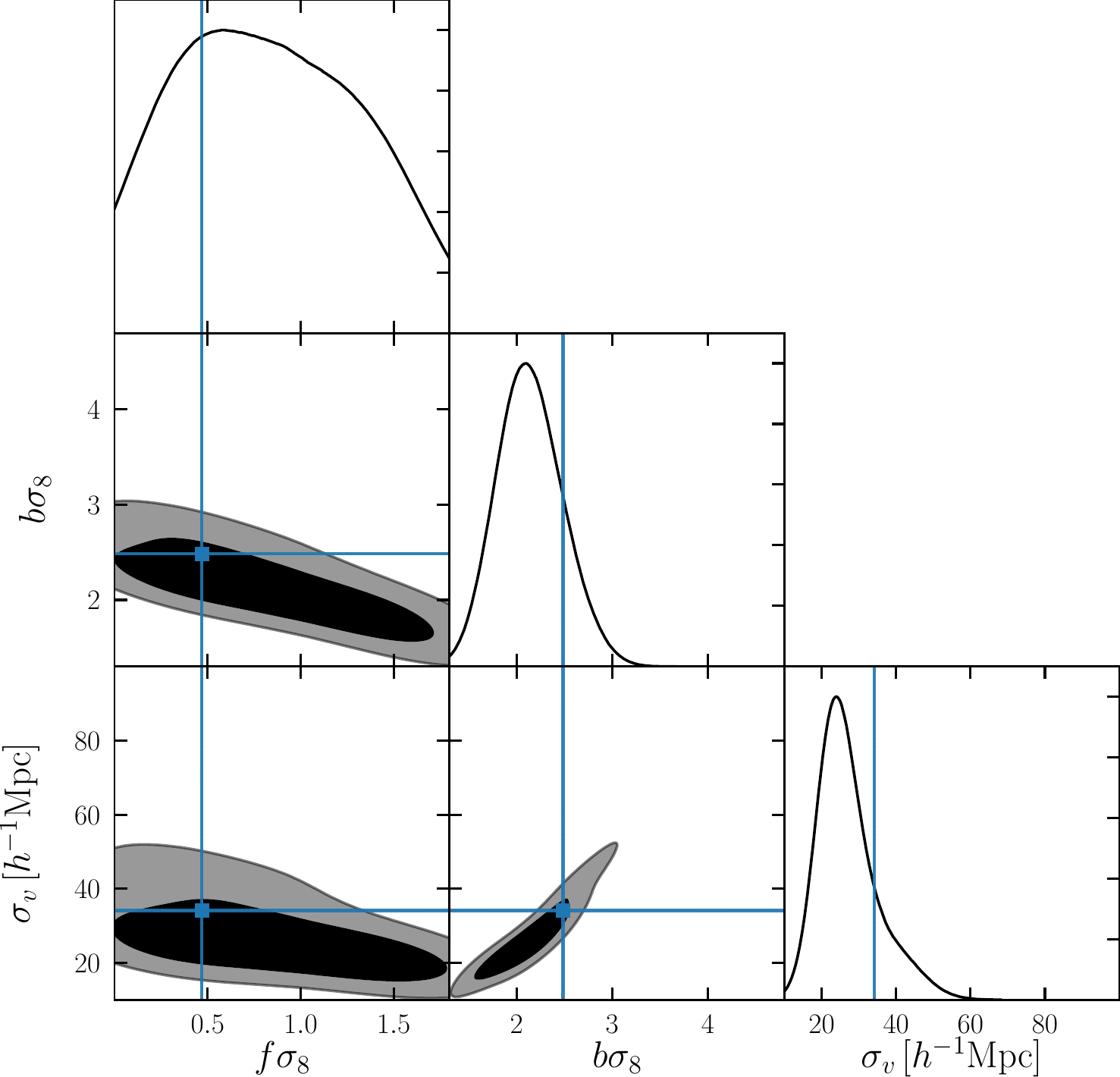}
\caption{Posterior of $f\sigma_8$, $b\sigma_8$, and $\sigma_\mathrm{v}$ from an individual DR1-like catalogue with $\sigma_\mathrm{z}=0.01(1+z)$. The contours correspond to the $68\%$ and $95\%$ credible regions and the truth is indicated in blue.}
\label{fig_R2}
\end{figure}

In Fig. \ref{fig_R2} we present a joint inference of $f\sigma_8$, $b\sigma_8$, and $\sigma_\mathrm{v}$ from an individual DR1-like catalogue with realistic redshift uncertainties. The inferred contours are consistent with the simulation truth. Possible degeneracies arising in higher dimensional parameter inferences, particularly when BAO are varied jointly, will be investigated in future analyses of later Euclid DRs, where the statistical constraining power will be substantially improved. The phenomenology of such degeneracies and potential mitigation strategies are discussed in \citet{karcher}.

\subsection{Contribution of the hexadecapole}\label{app:hexadecapole}

\begin{figure}
\centering
\includegraphics[width=0.45\textwidth]{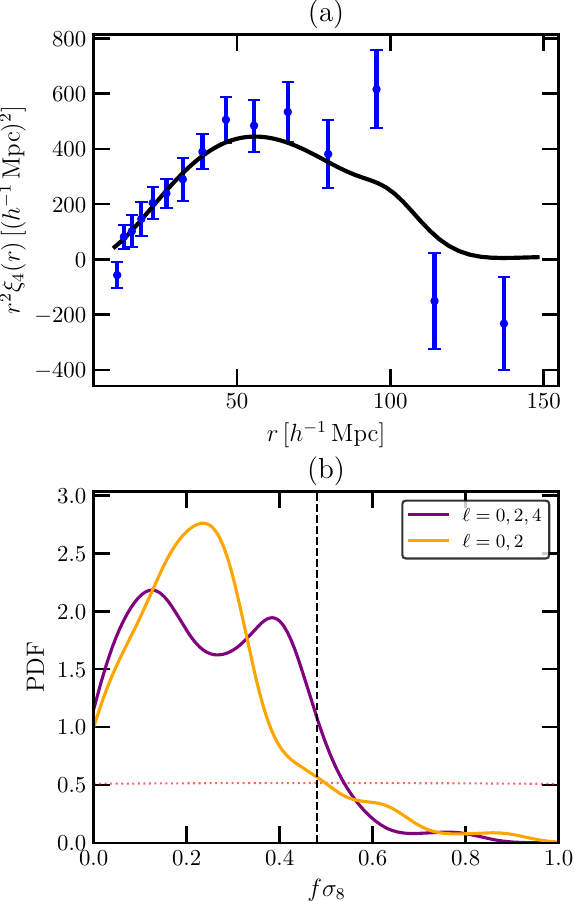}
\caption{(a) Hexadecapole as a function of scale for one random realisation in the $0.4<z<0.8$ bin for realistic redshift uncertainties (blue) compared to the true model prediction (black). (b) $f\sigma_8$ posteriors with (purple) and without (orange) the hexadecapole inferred by fitting the dispersion model to the mock data in panel (a). The prior is shown in red and the vertical dashed line indicates the truth in the simulation.}
\label{fig_9}
\end{figure}

For the investigation of the contribution of $\ell=4$, we compare the $f\sigma_8$ posteriors with and without including it in the inference. We work in the highest-S/N redshift bin as above, and consider $r_\mathrm{min,4}=r_\mathrm{min,0}=10$ $h^{-1}$Mpc, following the assumptions of \citet{karcher}. In Fig. \ref{fig_9} we present our results on the $f\sigma_8$ inference with and without the hexadecapole. As a simple scalar measure of information content, we compare the inverse posterior variances of $f\sigma_8$, and estimate the uncertainty on their ratio $I_{\ell=0,2,4}/I_{\ell=0,2}$ by bootstrap resampling the $f\sigma_8$ samples in each case. We find an information ratio of $1.04^{+0.12}_{-0.11}$, consistent with unity. Thus, at DR1-like S/N, including the hexadecapole does not yield appreciable constraints even in the highest-S/N redshift bin with realistic redshift uncertainties. In future DRs where the hexadecapole could be more informative, the grid search strategy we employ here will be computationally expensive for a similar analysis and therefore needs to be reconsidered. 

\subsection{Statistical requirements and conditions}\label{app:stat_req}

\begin{figure*}
\centering
\includegraphics[width=\textwidth]{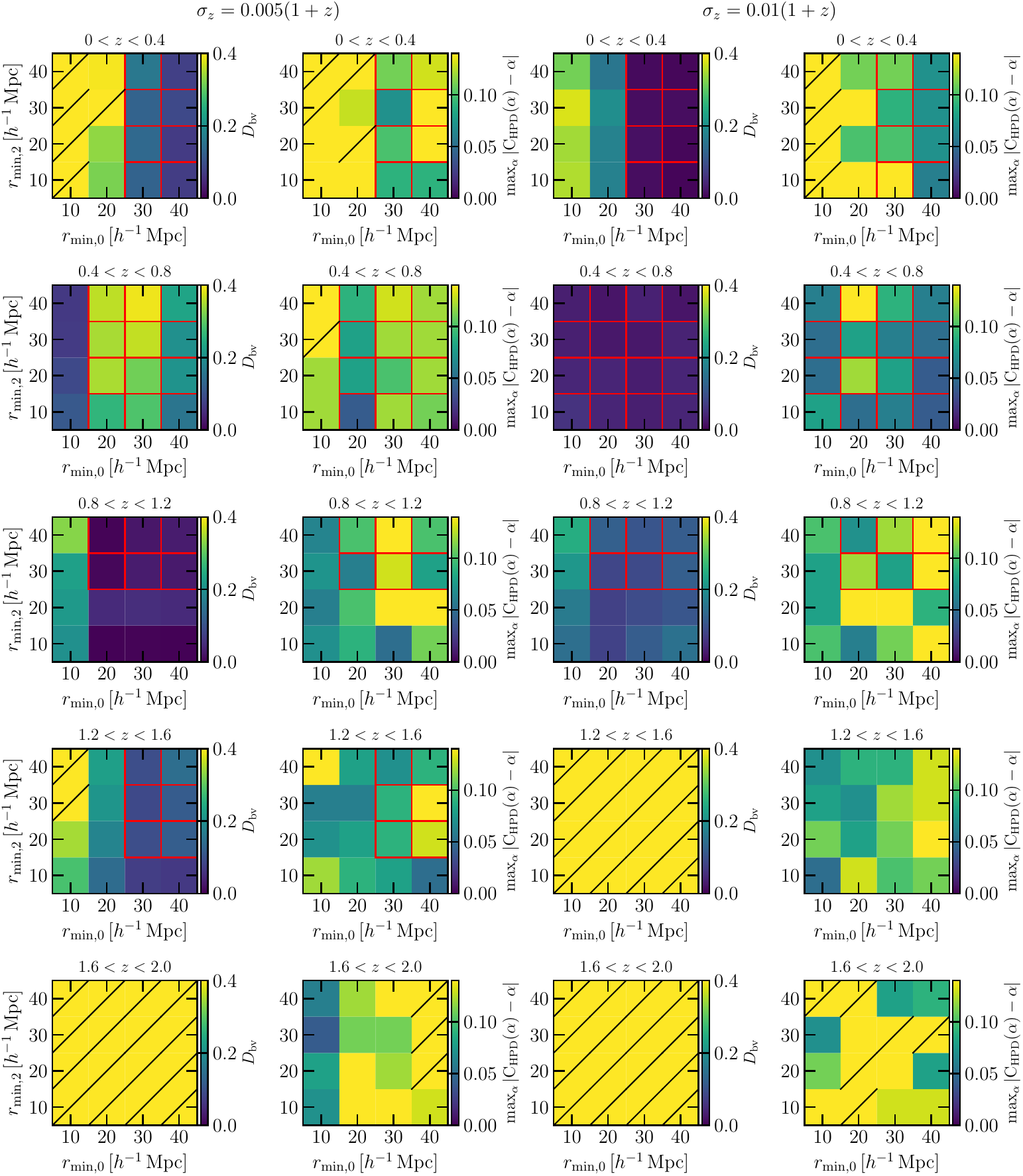}
\caption{Distance in the bias-variance plane (column 1, 3) and HPD interval (column 2, 4) on the ($r_\mathrm{min,0},r_\mathrm{min,2}$) plane, for optimistic and realistic redshift uncertainties, respectively, as a function of redshift. The black lines indicate the scales excluded by Eqs. \eqref{eq:D_bv} and \eqref{eq:hpd} due to failure of the estimator for quasi-uniform $f\sigma_8$ posteriors in a DR1-like setting. The red squares highlight the scales admitted after all conditions in Sect. \ref{sec:criteria} are applied, as shown in Table \ref{tab:results}.}
\label{fig_10}
\end{figure*}

\begin{table}[!ht]
\caption{\label{tab:redchi}Mean reduced $\chi^2$ values under the conservative validation criteria.}
\centering
\renewcommand{\arraystretch}{1.2}
\begin{tabular}{ccc}
\hline\hline
Redshift bin & $\sigma_0 = 0.005$ & $\sigma_0 = 0.01$ \\
\hline
$0.0 < z < 0.4$ & $0.97 \pm 0.35$ & $1.02 \pm 0.40$ \\
$0.4 < z < 0.8$ & $1.13 \pm 0.36$ & $1.06 \pm 0.40$ \\
$0.8 < z < 1.2$ & $1.04 \pm 0.34$ & $1.09 \pm 0.36$ \\
$1.2 < z < 1.6$ & $1.21 \pm 0.44$ & -- \\
$1.6 < z < 2.0$ & -- & -- \\
\hline
\end{tabular}
\tablefoot{The values correspond to a DR1-like \textit{Euclid} cluster sample, assuming the \citetalias{tinker} halo-bias prescription and the conservative statistical tests. Dashes indicate bins excluded because the estimator is biased for quasi-uniform $f\sigma_8$ posteriors.}
\end{table}

In Fig.~\ref{fig_10} we show the results of the calibration tests based on the distance in the bias–variance plane and the HPD-coverage statistic, defined in Eqs.~\eqref{eq:D_bv} and \eqref{eq:hpd}. For optimistic redshift uncertainties, and at low redshift, both $D_\mathrm{bv}$ and $\max_\alpha |\tens{C}_\mathrm{HPD}(\alpha)-\alpha|$ exclude a sizable region of small-scale cuts. At higher redshift, $D_\mathrm{bv}$ becomes a biased estimator. For more realistic redshift uncertainties, $\max_\alpha |\tens{C}_\mathrm{HPD}(\alpha)-\alpha|$ retains some constraining power in the lowest-redshift bin. The relative behaviour of $D_\mathrm{bv}$ and $\max_\alpha |\tens{C}_\mathrm{HPD}(\alpha)-\alpha|$ at high redshift remains similar to that seen in the optimistic-redshift case owing to the bias of the estimators. The behaviour of $\max_\alpha |\tens{C}_\mathrm{HPD}(\alpha)-\alpha|$ in the $1.2<z<1.6$ bin for $\sigma_\mathrm{z}=0.005(1+z)$ illustrates why we propose a monotonicity condition on the $(r_{\min,0},r_{\min,2})$ plane to avoid interpreting statistical noise fluctuations as signal. Although the pair $(r_{\min,0},r_{\min,2})=(10,10)\,h^{-1}\,\mathrm{Mpc}$ would formally satisfy the $D_\mathrm{bv}$ criterion, neighbouring pairs with larger $r_{\min,0}$ fail the $D_\mathrm{bv}$ test, indicating that the apparent success at $(10,10)$ is not robust. Under our monotonicity requirement, such isolated “passing” pairs are therefore discarded. A complementary example of the elimination of unstable scales is seen in the $D_\mathrm{bv}$ plane for $0<z<0.4$ and $\sigma_\mathrm{z}=0.01(1+z)$: although the pair $(r_{\min,0},r_{\min,2})=(20,40)\,h^{-1}\,\mathrm{Mpc}$ satisfies both $D_\mathrm{bv}\lesssim0.4$ and $\max_\alpha |\tens{C}_\mathrm{HPD}(\alpha)-\alpha| \lesssim 0.14$, $D_\mathrm{bv}$ exhibits a steep, sudden increase for $r_{\min,0}\leq 20\,h^{-1}\,\mathrm{Mpc}$, which we interpret as the onset of model mis-specification at smaller scales under these much stricter requirements. Our monotonicity requirement therefore leads us to discard this scale pair as well. Finally, for the scales surviving the above criteria, as discussed in Sect. \ref{sec:criteria}, we further use the mean of the reduced $\chi^2$ distribution as an additional test based on the fitted 2PCF rather than on the inferred $f\sigma_8$. In Table~\ref{tab:redchi} we report the mean $\chi^2_\mathrm{red}$ values for the scales reported for the conservative tests in Sect. \ref{sec:results}, showing consistency with unity.

\section{Non-linear clustering models}\label{app:nl_clustering}

There is a wide range of non-linear models to be considered for the clustering analyses of upcoming DRs, including: the velocity difference generator \citep{2025PhRvD.112f3532E}, convolution Lagrangian perturbation theory \citep{2013MNRAS.429.1674C} and effective field theory (EFT) \citep{2016JCAP...12..007V}, which were considered in the galaxy clustering analysis by \citet{karcher}. Additionally, the full 1-loop theory of redshift-space clustering in the EFT of the large-scale structure has been implemented by \citet{2020JCAP...07..062C,2020PhRvD.102f3533C,2021JCAP...01..006D}, and \citet{2022JCAP...11..038N}. Even though in a DR1-like setting the dispersion model passes the DR1-like validation criteria considered in the main analysis, in this appendix we present our prediction with a model closer to the state-of-the-art and compare its predictions against the 2PCF with uncertainties rescaled to DR2- and DR3-like footprints to explore whether the fit at high redshifts can be improved. For this purpose, we use the \texttt{velocileptors}\footnote{\href{https://github.com/sfschen/velocileptors}{github.com/sfschen/velocileptors}} model that employs one-loop perturbation theory with effective field theory terms. Here, we do not perform a full EFT likelihood analysis with all nuisance parameters marginalised; instead, we fit a linear and a quadratic bias term against the mean DR1-like 2PCF with $\sigma_\mathrm{z}=0.01(1+z)$. This comparison is intended as a diagnostic of whether a more advanced clustering model improves the residuals relative to the baseline 2PCF.

\begin{figure*}
\centering
\includegraphics[width=\textwidth]{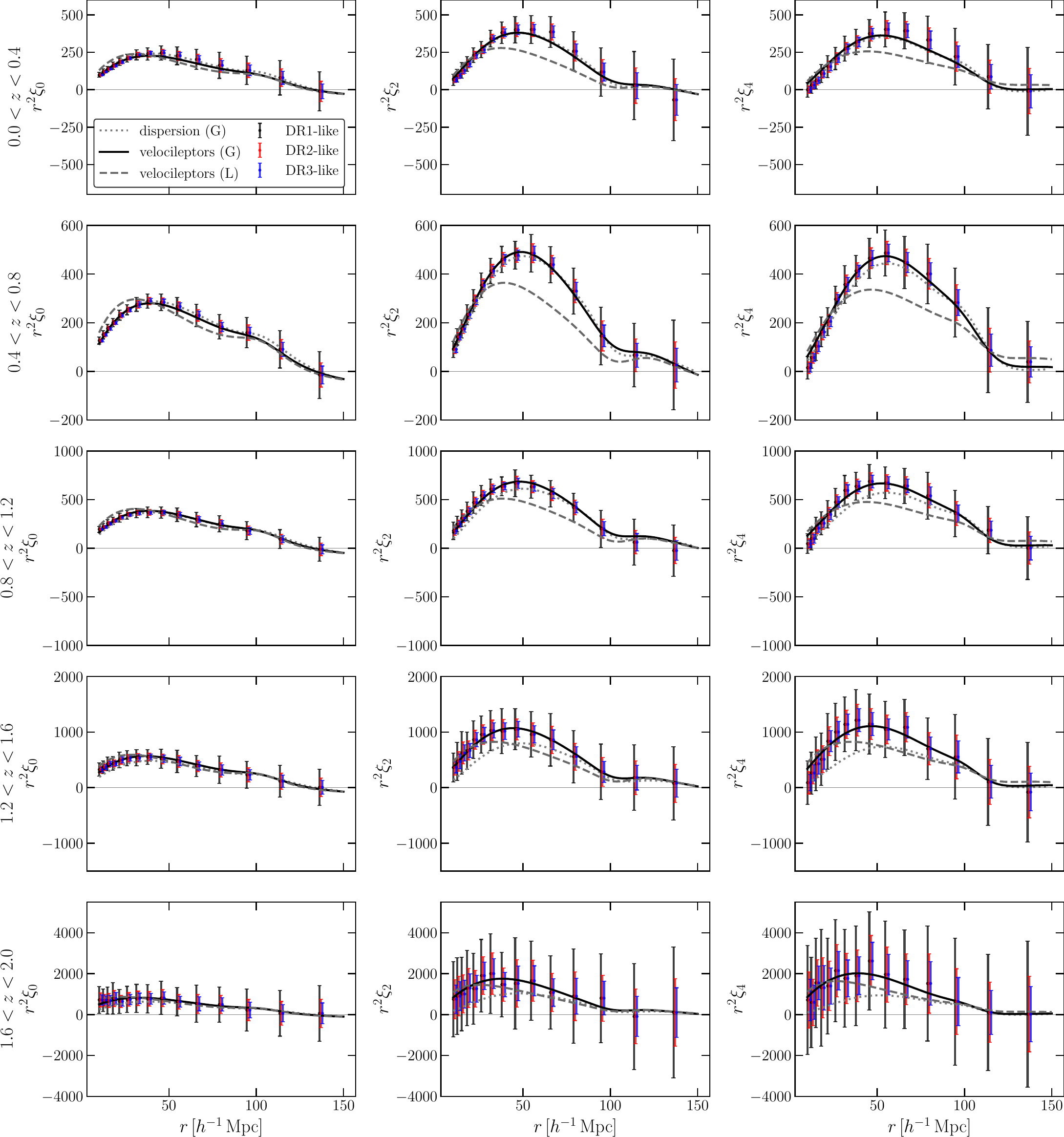}
\caption{The \texttt{velocileptors} model (black solid line) fitted against the mean 2PCF in the more realistic DR1-like setting compared to the simpler dispersion model (grey dotted line). In red and blue, we show the error bars rescaled to the DR2 and DR3 footprint. The fitted model is presented with Gaussian (G) and Lorentzian (L, grey dashed line) damping.}
\label{fig_R1}
\end{figure*}

We show the results in Fig. \ref{fig_R1}, overlaying the 2PCF error bars obtained by rescaling the DR1-like covariance to the DR2-like and DR3-like survey areas. The comparison indicates that a more flexible model can improve the high-redshift fit. Replacing the Gaussian damping by a Lorentzian kernel does not remove the high-redshift residuals or provide a qualitatively better description of the mean multipoles. Since the photometric redshift uncertainties in the mocks are Gaussian by construction, this test suggests that the residuals are not primarily caused by choosing an incorrect damping kernel.

\begin{figure}
\centering
\includegraphics[width=0.47\textwidth]{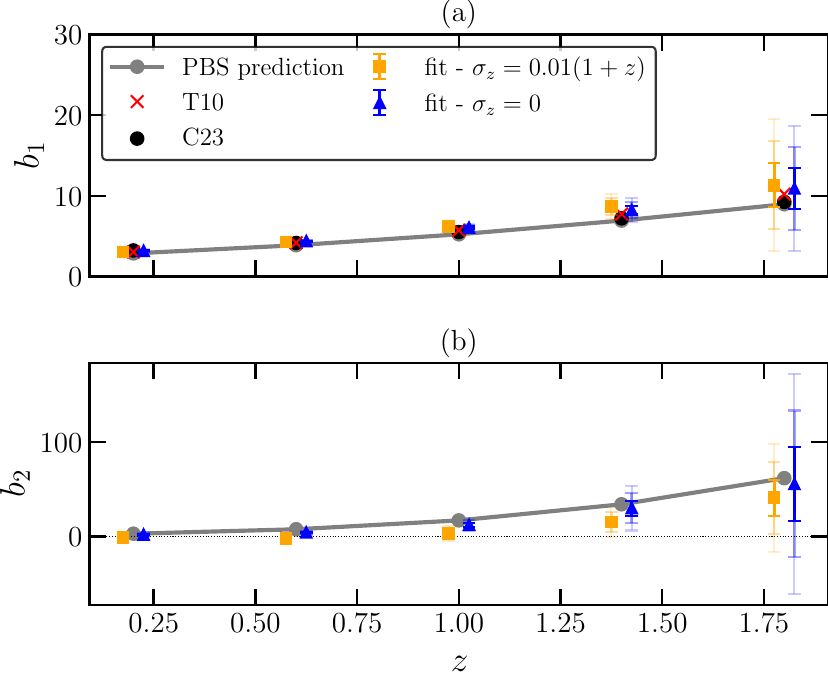}
\caption{(a) Linear and (b) quadratic bias terms of the \texttt{velocileptors} model fitted against the mean 2PCF with DR1-like redshift uncertainties (orange) and without redshift uncertainties (blue) as a function of redshift. The grey line represents the PBS prediction, whereas the red cross and black dot represent the \citetalias{tinker} and \citetalias{castro23} predictions, respectively. A small horizontal offset is applied to each point to facilitate comparison.}
\label{fig_R3}
\end{figure}

We use the peak-background split (PBS) formalism to derive predictions for the fitted linear, $b_1$, and quadratic, $b_2$, bias terms and compare them to the values fitted with the \texttt{velocileptors} model in Fig. \ref{fig_R3}. The upper panel shows that the fitted linear bias follows the expected increase with redshift for the mass-selected cluster sample. The \citetalias{castro23} prescription is close to the PBS expectation, as expected since the former is a calibrated correction to the latter. The \citetalias{tinker} values differ at the several-percent level. The fitted values are systematically higher than the PBS prediction, although they follow the same redshift trend. The values obtained with and without photometric redshift uncertainties are mutually consistent within their uncertainties. This indicates that the offset in the fitted linear amplitude is not primarily caused by the photometric redshift smearing, but should instead be interpreted as part of the effective bias calibration of the restricted two-point model over the fitted range of scales.

The lower panel shows that the PBS prediction for $b_2$ rises rapidly with redshift. The fitted $b_2$ values show the same overall tendency towards larger values at higher redshift, but lie below the local PBS prediction, particularly in the high-redshift bins. Further, the fitted $b_2$ is systematically lower when DR1-like redshift uncertainties are included. Therefore, the fitted $b_2$ should be interpreted as an effective coefficient, not as the physical local quadratic bias, since it may also absorb contributions from higher order bias terms, which cannot be safely constrained with our mock 2PCFs.

\end{appendix}

\end{document}